\documentclass[11pt]{article}
\usepackage{amsmath,amsthm,amsfonts,amscd,eucal,latexsym,amssymb}
\usepackage{epsfig}  
\usepackage[latin1]{inputenc}
\usepackage{verbatim} 
\input xy
\xyoption{all}
\def\bC{{\mathbb C}}           

\def\bR{{\mathbb R}}

\def\bZ{{\mathbb Z}}

\def\beq{\begin{eqnarray}}
\def\eeq{\end{eqnarray}}

\newtheorem{proposition}{Proposition}[section]
\newtheorem{lemma}{Lemma}[section]
\newtheorem{definition}{Definition}[section]

\newcommand{\se}[1]{\section{#1}}

\def\vsp{\vspace{0.2cm}}
\def\vspp{\vspace{0.1cm}}

\def\sse #1 {\vsp\ifhmode{\par}\fi\refstepcounter{subsection}
  \noindent {\bf\thesubsection}. {\em #1}.\quad
  \addcontentsline{toc}{subsection}{\protect\numberline{\thesubsection} #1}%
  }

\def\ssb #1 {\vsp\ifhmode{\par}\fi\refstepcounter{subsection}
  \noindent {\bf\thesubsection.} {\bf #1.}\quad
  \addcontentsline{toc}{subsection}{\protect\numberline{\thesubsection} #1}%
  }

\def\ssa #1 {\ifhmode{\par}\fi\refstepcounter{subsection}
  \noindent {\bf\thesubsection.} {\bf #1.}\quad
  \addcontentsline{toc}{subsection}{\protect\numberline{\thesubsection} #1}%
  }

\def\remark #1 {\vsp\vspp\ifhmode{\par}\fi\noindent\noindent {\bf Remark.} {#1}\vsp\vspp\par}
\def\remarks #1 {\vsp\vspp\ifhmode{\par}\fi\noindent\noindent {\bf Remarks.} {#1}\vsp\vspp\par}

\begin{document}

\hfill{\sl February 2011}
\par
\bigskip
\par
\rm


\par
\bigskip
\LARGE
\noindent
{\begin{center}\bf Remarks on the Reeh-Schlieder property for higher spin free fields on curved spacetimes\end{center}}
\bigskip
\bigskip
\par
\rm
\normalsize


\noindent {\bf Claudio Dappiaggi$^{1,a}$}\\
\par
\small
\noindent $^1$
Dipartimento di Fisica Nucleare e Teorica, Universit\`a degli Studi di Pavia \& INFN Sezione di Pavia, Via Bassi 6, I-27100 Pavia (PV), Italia.\\
\noindent $^a$  claudio.dappiaggi@pv.infn.it\\
 \normalsize

\par

\rm\normalsize
\noindent {\small Version of \today}

\rm\normalsize


\par
\bigskip

\noindent
\small
{\bf Abstract}.
The existence of states enjoying a weak form of the Reeh-Schlieder property has been recently established on curved backgrounds and in the framework of locally covariant quantum field theory. Since only the example of a real scalar field has been discussed, we extend the analysis to the case of massive and massless free fields either of spin $\frac{1}{2}$ or of spin $1$. In the process, it is also shown that both the vector potential and the Proca field can be described as a locally covariant quantum field theory.
\normalsize

\se{Introduction}
One of the main results in the algebraic approach to quantum field theory is the so-called {\em Reeh-Schlieder} theorem, which asserts that, in the framework of a real scalar field theory on Minkowski spacetime, the set of vectors generated from the vacuum by the polynomial algebra of any open region is dense in the underlying Hilbert space -- see theorem 5.3.1 in \cite{Haag}. From a physical point of view this statement basically warns us that making precise the qualitative notion of localized states can be trickier than it appears at first glance. This property was shown, moreover, to hold true for a broader class of free fields and on a larger class of backgrounds in \cite{Strohmaier} where the existence of an algebraic state enjoying the Reeh-Schlieder property was proven for scalar, Dirac and Proca fields on a four dimensional static manifold. This result prompted a further analysis in \cite{Sanders2}: In the framework of locally covariant field theories it was shown that the construction of a state enjoying the Reeh-Schlieder property on an ultrastatic spacetime suffices to claim the existence of a state which enjoys the very same property at least for a suitable region on an arbitrary globally hyperbolic spacetime. This is achieved by a careful use of a deformation argument first introduced in \cite{FNW} and often used in the literature to prove the existence of states enjoying the so-called Hadamard condition \cite{Sahlmann2}. Furthermore, in \cite{Sanders2} it was also shown that a free real scalar field falls in the class of theories for which the mentioned result can be applied. 

The goal of this paper is to prove that the existence of a state enjoying at least on a suitable region the Reeh-Schlieder property is not restricted to the example of a scalar field theory, but it can be extended to higher spin field theories, most notably the Dirac field, the Proca field and the vector potential. The reason for this investigation originated not only from the desire of extending the axiomatic results on Minkowski background to an arbitrary globally hyperbolic spacetime, but it is also connected to some recent results in constructive algebraic quantum field theory. In particular, in \cite{DLM} it was shown that warped convolutions, a deformation technique first developed for quantum field theories on Minkowski space and closely related to Rieffel's deformation \cite{BLS}, can be applied also on a rather broad class of curved backgrounds. Although the whole procedure is carried in a model-independent, operator-algebraic framework, and although the deformed models share many structural properties with deformations of quantum field theories on a flat background, a natural and yet very difficult question to answer is whether the deformed model is actually not equivalent to the original one. Although a general answer is still not within reach, thanks to the results in \cite{BLS}, it was shown that this is indeed the case for a free Dirac field living on a Friedmann-Robertson-Walker spacetime with flat spatial section. The key point of the argument is the existence of a state for the undeformed theory which enjoys the Reeh-Schlieder property for a suitable open region, a fact which cannot hold true after the deformation.

Hence, on account of this example, it is even more important to answer the question whether higher spin free field theories posses a state which enjoys at least locally the Reeh-Schlieder property. While the case of a Dirac field is rather simple since it was already proven in \cite{Sanders3} that it can be described as a locally covariant quantum field theory, a similar analysis for the Proca field and for the vector potential seems not to be available in the literature. 

Hence, we organize the paper as follows: In section 2.1, we introduce with the language of bundles the kinematical configurations for spin $\frac{1}{2}$ and spin $1$ fields on an arbitrary globally hyperbolic spacetime $(M,g)$.
In section 2.2 the classical dynamics are instead thoroughly discussed and the associated algebras of observables are defined. In section 2.3 we recast the previous analyses in the language of general local covariance and it is proven that both the vector potential and the Proca field fit in this framework. Section 3 is instead entirely devoted to the recollection of the deformation argument; this leads to the identification of a state enjoying at least on a suitable region the Reeh-Schlieder property and to the proof that such a state exists also for higher spin free field theories.

\section{Higher spin field: classical dynamics and field algebra}\label{CFT}

The determination of a classical dynamical system is essentially ruled by two key steps: the construction of suitable kinematical configurations and the identification of a distinguished subset ruled by the underlying dynamics. If one focuses only on non-interacting theories constructed on the flat Minkowski background, there exists a well-established procedure, first introduced by Wigner, which completely and unambiguously classifies the class of possible free fields and their equation of motion out of the unitary and irreducible representations of the Poincar\'e group.

On a manifold with a non trivial Lorentzian metric the situation is less crystal clear due to the potential absence of any isometry and hence of any Wigner-like analysis. In such a framework even the construction of a free field theory requires a more careful analysis and it is ultimately affected by some arbitrary choices. Even under the reasonable requirement that the generalization on curved backgrounds of a free field should coincide with the Minkowski one in the limit when the metric becomes the flat one, already the definition of suitable kinematically allowed configurations is a rather subtle affair. The first problem one faces is the necessity to find a suitable notion of a ``field with spin $j$'', $j$ being any integer or semi-integer number. Although for scalar and for spin $1$ fields, one can easily consider smooth function and $1$-forms respectively, the case where $j=\frac{1}{2}$ is already less transparent. In this case one is forced to develop a characterization of the spin of a field out of geometric structures which are meaningful also on a large class of manifolds with a non-trivial geometry. An elegant solution to this potential difficulty can be found combining the language of fibre bundles with the existence of a natural notion of Lorentz group connected to the frames defined at each point of a Lorentzian background. The first goal of this section will be to recapitulate such an approach for the non-trivial scenario of Dirac field and later we shall extend it to the case of spin $1$. Since this last case was never presented in this language, we feel worth to devote a few lines to close this small gap.

\subsection{The bundle structures and the kinematical configurations}

As a first step we shall introduce the kinematically allowed configurations for spin $\frac{1}{2}$ and spin $1$ fields  leaving the discussion of the dynamical ones to the next section. Due to a few intrinsic differences we shall always discuss the two cases separately. Furthermore, since there exists a recent and extensive literature on spinors, particularly \cite{DHP, Sanders, Sanders2, Verch} but also \cite{Dimock}, we will follow these references to which we refer for a more extensive analysis. Before entering into the details, we fix once and for all our notion of spacetime:

{\definition A spacetime is a four-dimensional differentiable, connected, Hausdorff manifold $M$ endowed with a smooth Lorentzian metric $g$ whose signature is chosen as $(-,+,+,+)$. Furthermore, if $M$ is oriented, time oriented and it possesses a Cauchy surface, that is a closed achronal set $\Sigma$ whose domain of dependence coincides with $M$, then $(M,g)$ is called globally hyperbolic.  Henceforth we shall only consider these manifolds.}\\

\vskip .3cm

\noindent{\em The kinematics of Dirac fields:} As we have already mentioned above, the notion of Dirac field requires first of all a suitable definition of spin and we shall now recall how this can be inferred from the geometric structures available on a generic spacetime.  

{\definition The Lorentz {\bf frame bundle} of $M$ is $LM=LM[SO_0(3,1),\pi_L,M]$ where $SO_0(3,1)$ is the component connected to the identity of the Lorentz group while $\pi_L:LM\to M$. This structure comes together with the free and transitive right action $R_L:SO_0(3,1)\times LM\to LM$ such that $R_L(\Lambda,[\Lambda',p])=[\Lambda\Lambda',p]$ for all $\Lambda\in SO_0(3,1)$ and for all $[\Lambda',p]\in LM$.}\\

Unfortunately, in order to define a suitable notion of spin structure, we need to go one step beyond the above definition and particularly we shall exploit that $SL(2,\bC)$ is the double cover of $SO_0(3,1)$. Hence,

{\definition\label{spinstr} We call {\bf spin structure} of a spacetime $(M,g)$, the following two data:
\begin{itemize}
\item the spin bundle $SM\doteq SM[SL(2,\bC),\pi_S,M]$ with $\pi_S:SM\to M$ and with the fibrewise right action $R_S$ of $SL(2,\bC)$,
\item a smooth bundle morphism $\rho:SM\to LM$ such that $\pi_L\circ\rho=\pi_S$ and $\rho\circ R_S(\tilde\Lambda)=R_L(\Lambda)\circ\rho$ where $\Lambda=\Pi(\tilde\Lambda)$, $\Pi$ being the canonical surjective group homomorphism from $SL(2,\bC)$ to $SO_0(3,1)$.
\end{itemize}}   
We recall that a spacetime $(M,g)$ admits a spin structure if and only if the second de Rham cohomology group with $\bZ_2$ coefficients $H^2(M,\bZ_2)=\{0\}$ and that the number of non equivalent ones coincides with that of the equivalence classes in $H^1(M,\bZ_2)$. Notice that in this paper all cohomolgies will be of de Rham type and hence we will not specify it again. Furthermore it is important to stress that, as proven by Geroch \cite{Geroch, Geroch2}, all globally hyperbolic spacetimes have a trivial second cohomology group with $\bZ_2$-coefficients. Hence, in the cases we consider, the existence of at least a spin structure is always guaranteed. We can thus proceed as follows 

{\definition\label{DS} The {\bf Dirac (spinor) bundle} of a four dimensional globally hyperbolic spacetime $M$ is the associated vector bundle $DM\doteq SM\times_T\bC^4$ where $T\doteq D^{\left(\frac{1}{2},0\right)}\oplus D^{\left(0,\frac{1}{2}\right)}$ is a $SL(2,\bC)$-representation. Hence $DM$ is the set of equivalence classes $[(p,z)]$ where $p\in SM$, $z\in\bC^4$ and $(p,z)\sim (p',z')$ if and only if there exists $\Lambda\in SL(2,\bC)$ such that $R_S(\Lambda)p=p'$ while $T(\Lambda^{-1})z=z'$. Hence $DM\equiv DM[\bC^4,\pi_D,M]$ is a fibre bundle over $M$ with $\bC^4$ as typical fibre. The projection map is inherited from $SM$, namely for every $[(p,z)]\in DM$, we set $\pi_D\left([(p,z)]\right)\doteq\pi_S(p)$. Furthermore, if we endow the typical fibre with the standard non-degenerate inner product, we can construct the {\bf dual Dirac bundle} $D^*M$ as the $\bC^{4*}$-bundle over $M$ where we require that the points $(p_1,z^*_1)$ and $(R_S(\Lambda)(p),z^*_1T(\Lambda))$ are equivalent. Here $*$ is the adjoint with respect to the inner product and thus elements of $\left(\bC^4\right)^*$ are meant as row vectors. Consequently, the dual pairing between $\bC^4$ and $\left(\bC^4\right)^*$ extends to a well-defined fibrewise dual pairing between $DM$ and $D^*M$.}\\

\noindent We can now introduce the key object of a spin $\frac{1}{2}$ field theory: 

{\definition\label{spinor} A {\bf spinor field} is a smooth global section of the Dirac bundle, namely $\psi\in\Gamma(DM)$. The space $\Gamma(DM)\doteq C^\infty(M,DM)$ is naturally endowed with the topology induced by the family of seminorms
$$\parallel\!\psi\!\parallel_{n,K}\doteq\sup\{|\psi^{(n)}(x)|,\; x\in K\},\qquad\psi\in\Gamma(DM)$$
where $(n)$ denotes the derivative of $n$-th order whereas $K$ is an arbitrary compact set. Equivalently, we call a {\bf cospinor field} a section $\psi'\in\Gamma(D^*M)$, the latter space being also endowed with the same topology as $\Gamma(DM)$.}\\

Notice that, together with smooth sections, it is important to consider also the space of {\em smooth sections with compact support}, $\mathcal{D}(DM)=C^\infty_0(M,DM)$ equipped with the following topology: A sequence $f_k\in\mathcal{D}(DM)$, $k\in\mathbb{N}$ converges to $f\in\mathcal{D}(DM)$ if all $f_k$ and $f$ are supported in a compact subset $K\subset M$ and all derivatives of $f_k$ converge to the ones of $f$ uniformly in $K$. A similar definition holds for $\mathcal{D}(D^*M)=C^\infty_0(M,D^*M)$ and furthermore there exists a global pairing between $\Gamma(DM)$ and $\mathcal{D}(D^*M)$ (as well as between $\Gamma(D^*M)$ and $\mathcal{D}(DM)$) by integrating the local pairing induced by the inner product on $\bC^4$:
$$\langle\psi,f\rangle\doteq\int\limits_M d\mu(x)\psi(x)(f(x)).\qquad\forall\psi\in\Gamma(DM),\;\textrm{and}\;\forall f\in \mathcal{D}(D^*M)$$

To summarize, on a globally hyperbolic spacetime, a kinematically allowed configuration of a spinor is nothing but a smooth section of an associated bundle constructed out of the same representation of $SL(2,\bC)$ which is used on Minkowski spacetime to characterize Dirac fields. Within this respect we can claim that $\psi$ as in definition \ref{spinor} has spin $\frac{1}{2}$. The generalization of this picture to the real or complex scalar field is rather straightforward, but it is certainly desirable not be limited to these cases, but, quite surprisingly, already spin $1$ fields seem not have been ever treated in this language. The usual paradigm to consider in this scenario sections of the cotangent bundle of the underlying spacetime is of course flawless, yet we find worth to supplement it with a complementary approach which justifies also on curved backgrounds the notion of spin $1$ fields. 

\vskip .3cm

\noindent{\em The kinematics of spin $1$ fields:} Let us briefly mention that, on Minkowski spacetime, a spin $1$ field, be it the Proca field or the vector potential, is real and unambiguously characterized by its transformation under a suitable unitary and irreducible representation of the Poincar\'e group induced from the $D^{\left(\frac{1}{2},\frac{1}{2}\right)}$-representation of $SL(2,\bC)$. Furthermore the reality condition for this field entails that such a representation boils down to the fundamental one of $SO_0(3,1)$. Therefore, in analogy with the Dirac case, but avoiding the notion of spin structure since not strictly necessary, we can proceed as follows:

{\definition\label{Proca} The {\bf (real) Proca bundle} of a spacetime $(M,g)$ is the associated bundle $PM\doteq LM\times_\Lambda\bR^4$ where $PM$ is the set of equivalence classes $[(p,v^\mu)]$, $p\in LM$ and $v^\mu\in\bR^4$, where $(p,v^\mu)\sim (p',v'^\mu)$ if and only if there exists $\Lambda\in SO_0(3,1)$ with $p'=R_L(\Lambda)p$ and $v'^\mu=\Lambda^\mu_\nu v^\nu$, where $\Lambda^\mu_\nu$ stands for the real fundamental representation of $SO_0(3,1)$ on $\bR^4$. In other words $PM=PM[\bR^4,\pi_P,M]$ where $\pi_P(p,v^\mu)\doteq\pi_L(p)$. A Proca field or a vector potential is $A\in\Gamma(PM)$, all the sections being endowed with the topology induced by the seminorms
$$\parallel\!A\!\parallel_{n,K}\doteq\sup\{|A^{(n)}(x)|,\; x\in K\},\qquad A\in\Gamma(PM)$$
where $(n)$ denotes the derivative of $n$-th order whereas $K$ is an arbitrary compact set.}\\

We stress that, in complete analogy with definition \ref{DS}, we could have started from an associated bundle constructed out of the $D^{\left(\frac{1}{2},\frac{1}{2}\right)}$-representation of $SL(2,\bC)$, considering ultimately only the real smooth sections. This turns out to be completely equivalent to definition \ref{Proca} and hence it is justified to call $A\in\Gamma(PM)$ a field of spin $1$.  

In order to see the connection with the standard picture of a Proca field or of a vector potential, we notice that $SO_0(3,1)$, the structure group, acts transitively on each fibre $\pi^{-1}(x)\sim\bR^4$ for all $x\in M$. Hence, we are free to solder $PM$ and to conclude that there exists a linear isomorphism $\Theta:TM\to PM$, which composed with the canonical metric-induced isomorphism $\Theta_g:T^*M\to TM$, yields the sought isomorphism $\widetilde\Theta:\Theta_g\circ\Theta:T^*M\to PM$. This entails that to any section $A\in\Gamma(PM)$ we can associate via pull-back a unique one form $\widetilde A\doteq\widetilde\Theta^*(A)\in\Gamma(T^*M)$. Henceforth, $\Theta^*$ will be left implicit and $A$ will always refer to a section of $T^*M$. To conclude notice also that the definition of $PM$ is sometimes used directly as the one of the tangent bundle \cite{KNVOL1}. 

\subsection{The dynamical configurations and the associated algebras}
Since we have a full control of the kinematically allowed configurations for spin $\frac{1}{2}$ and spin $1$ fields on a globally hyperbolic curved spacetime, we can focus on the dynamics. We shall split our discussion in three parts covering Dirac fields, Proca fields and vector potentials respectively. Furthermore we will also introduce the algebra of observables of each of these theories since it is unambiguously determined out of the space of solutions of the equations of motion.

\vskip .2cm

\noindent{\em The dynamics and the algebra of fields of Dirac fields:} This scenario has been thoroughly discussed in the recent literature, see for example \cite{Dimock, DHP, Hack, Sanders} and we shall thus only recall the main ingredients and results since a careful analysis would require a paper on its own. Notice that in this section we shall adopt the notation and the nomenclature of \cite{DHP}. The dynamics for a spinor $\psi\in\Gamma(DM)$ and for a cospinor $\psi'\in\Gamma(D^*M)$ are ruled by the Dirac equation:
\beq\label{Diraceq}\left\{\begin{array}{l}
D\psi=(-\gamma^\mu\nabla_\mu+m\mathbb{I})\psi=0,\\
D'\psi'=(\gamma^\mu\nabla_\mu+m\mathbb{I})\psi'=0
\end{array}\right. ,
\eeq
where $m\geq 0$, $\mathbb{I}$ is the identity operator and where the $\gamma$-matrices are chosen as 
$$\gamma_0=\left(\begin{array}{cc} 
I_2 & 0\\
0 & I_2
\end{array}\right), \qquad \gamma_i=\left(\begin{array}{cc} 
0 & \sigma_i\\
-\sigma_i & 0
\end{array}\right),$$
$\sigma_i$ with $i=1,2,3$ being the standard Pauli matrices and $I_2$ the $2\times 2$ identity matrix. Furthermore $\Delta:\Gamma(DM)\to\Gamma(DM\otimes T^*M)$ is the standard spin connection -- see definition 2.9 in \cite{DHP}. As proven in \cite{Dimock, DHP}, the operator $D$ admits an advanced $(^-)$ and a retarded $(^+)$ fundamental solution $S^\pm:\mathcal{D}(DM)\to\Gamma(DM)$ fulfilling $DS^\pm=\textrm{id}=S^\pm D$, id being the identity map on the appropriate spaces. Furthermore $S^\pm$ enjoy the standard support property $\textrm{supp}(S^\pm(f))\subseteq J^\pm(\textrm{supp}(f))$  for all $f\in\mathcal{D}(DM)$. An identical result holds true for the cospinor and for the operator $D'$ which thus also admits a unique advanced and retarded fundamental solution $S^\pm_*:\mathcal{D}(D^*M)\to\Gamma(D^*M)$ enjoying the same properties as $S^\pm$. Till this point, spinors and cospinors have been considered as completely distinct objects although, as customary in Minkowski spacetime, it is possible to relate them via a suitable mapping. To this avail, we recall that there exists the {\em Dirac conjugation matrix}, that is the unique $\beta\in SL(4,\bC)$ such that (i) $\beta^*=\beta$, (ii) $\gamma^*_a=-\beta\gamma_a\beta^{-1}$ with $a=0,...,3$ and (iii) $i\beta n^a\gamma_a>0$, $n$ being timelike and future-directed. This object allows us to introduce the {\bf Dirac conjugation maps}:
\begin{gather*}
\cdot^\dagger:\Gamma(DM)\to\Gamma(D^*M),\qquad f^\dagger\doteq f^*\beta,\\
\cdot^\dagger:\Gamma(D^*M)\to\Gamma(DM),\qquad h^\dagger\doteq\beta^{-1}h^*,
\end{gather*} 
where $*$ denotes the adjoint with respect to the inner product on $\bC^4$. Notice both that $\beta$ is unique once a choice of the $\gamma$-matrices has been made and that, if we apply the Dirac conjugation maps twice consecutively, we obtain the identity. In particular in our scenario $\beta=-i\gamma_0$.

\vskip .2cm

\noindent As it holds true for bosonic field theories, the full control of the classical dynamics entails the possibility to introduce a natural algebra of observables. Within this respect, the first formulation of Dirac fields on curved backgrounds \cite{Dimock} treats ``particles" and ``antiparticles" as distinct objects although, since we are interested in the locally covariant properties of spinors, we shall follow a different approach first introduced in \cite{Araki} and later used also in \cite{DHP, Sanders}. This calls for considering spinors and cospinors as part of a single object; the building block of this idea is the direct sum of vector bundles $DM\oplus D^*M$ out of which we can define $\mathcal{D}\doteq\mathcal{D}(DM\oplus D^*M)$. It is the set of smooth and compactly supported sections endowed with the standard topology induced from that of $\mathcal{D}(DM)$ and of $\mathcal{D}(D^*M)$ as per definition \ref{spinor}. The Dirac conjugation maps above defined can be joined to form the new application $\Gamma:\mathcal{D}\to\mathcal{D}$ such that 
\beq\label{Gamma}
\Gamma(f\oplus h)\doteq h^\dagger\oplus f^\dagger,\qquad\forall f\oplus h\in\mathcal{D}.
\eeq
In order to define a suitable algebra we need a last datum, that is a sesquilinear form $(,):\mathcal{D}^2\to\bC$:
\beq\label{sesqui}
(f,h)\doteq -i\langle f^\dagger_1,Sh_1\rangle+i\langle S_*h_2,f^\dagger_2\rangle,
\eeq
where $f\doteq f_1\oplus f_2$, $h\doteq h_1\oplus h_2$ and $\langle,\rangle$ is the non-degenerate pairing between $\Gamma(DM)$ and $\mathcal{D}(D^*M)$ introduced in the previous section. 

{\definition\label{Diracalg} We call {\bf algebra of fields} of the Dirac field, the unital $*$-algebra $\mathcal{F}(M,g)$ generated by the identity and the abstract elements $B(f)$ with $f\in\mathcal{D}$. They satisfy the following defining relations:
\begin{itemize}
\item the map $f\mapsto B(f)$ is linear,
\item $B(Df_1\oplus D'f_2)=0$ for all $f_1\oplus f_2\in\mathcal{D}$, where $D$ and $D'$ are the operators defined in \eqref{Diraceq},
\item $B(\Gamma f)=B(f)^*$ for all $f\in\mathcal{D}$ and with $\Gamma$ defined as in \eqref{Gamma},
\item $\{B(f)^*,B(h)\}=B(f)^*B(h)+B(h)B(f)^*\doteq(f,h)$ for all $f,h\in\mathcal{D}$ and where $(,)$ is defined as in \eqref{sesqui}.
\end{itemize}}

We remark that the notion of spinor and cospinor can be recovered from the generators of the field algebra by a suitable choice of the test-functions, namely $\psi(h)\doteq B(0\oplus h)$ where $\psi^\dagger(f)= B(f\oplus 0)$. Furthermore, since $(,)$ is a sesquilinear form, we can consider the coset space $\mathcal{D}/\left(Ker(S\oplus S_*)\right)$ and complete it to a Hilbert space $\mathcal{H}$ with respect to $(,)$. As a by-product the pair $(\mathcal{H},\Gamma)$ allows the extension of $\mathcal{F}(M,g)$ to a C$^*$-algebra, $\mathfrak{F}(M,g)$ whose elements are bounded operators on $\mathcal{H}$ itself. Yet it is clear that not all elements of $\mathfrak{F}(M,g)$ can be considered as genuine observables due to the anticommuting nature of spinors and cospinors. Hence, as a first step, we restrict our attention to $\mathcal{F}_{even}(M,g)$, the even subalgebra of $\mathfrak{F}(M,g)$ whose elements are invariant under the map $B(f)\mapsto -B(f)$. As shown in proposition 3.1 of \cite{DHP}, this suffices to guarantee that elements, whose supports are spacelike separated, commute. This is still not sufficient since we have also to ensure that all elements are ``well behaving'' under the action of an element of $SL(2,\bC)$. Hence, we shall consider as {\em algebra of observables} the set $\mathcal{A}(M,g)\subset\mathcal{F}_{even}(M,g)$ whose elements $A\doteq\sum_n B(f_{n,1})...B(f_{n,2k_n})$ are (pointwise) invariant under a particular ``action'' $\widetilde L_z(\Lambda)$ of any $\Lambda\in Spin_0(3,1)$; given an arbitrary but fixed $z\in\bC^4$ and any $p\in SM$, we first define $\widetilde L_z(\Lambda)$ on $[(p,z)]\in DM$ and $[(p,z^*)]\in D^*M$ as $$\widetilde L_z(\Lambda)\left([(p,z)]\right)\doteq [(p,T(\Lambda)z)], \quad \widetilde L_z(\Lambda)\left([(p,z^*)]\right)\doteq [(p,z^*T(\Lambda^{-1}))],$$ 
where $T$ is the same representation of $SL(2,\mathbb{C})$ introduced in definition \ref{spinor}. $\widetilde L_z(\Lambda)$ can then be straightforwardly extended to $DM\oplus D^*M$, subsequently to arbitrary outer tensor products of the latter, and finally to the test sections $f_{n,1}\otimes \cdots \otimes f_{n,2k_n}$ determining $A$. Since $\widetilde L_z(\Lambda)$ depends on $z$, it is of course not a well-defined action in the strict sense, {\it cf.}, the footnote on page 74 in \cite{Sanders}.  
 
\vskip .2cm

{\noindent\em The dynamics and the Weyl algebra of Proca fields:} We turn our attention to the case of spin $1$ fields and due to some sharp differences we discuss separately the massless and the massive case. We shall start from the latter already considered in \cite{Furlani, Fewster}. We say that $A\in\Gamma(T^*M)$ is a dynamically allowed Proca field if
\beq\label{eom}
(\delta d+m^2)A=0,
\eeq
where $m^2>0$, while $d:\Omega^p(M)\to\Omega^{(p+1)}(M)$ is the exterior derivative, $\Omega^p$ being the space of smooth real-valued $p$-forms. At the same time $\delta:\Omega^p(M)\to\Omega^{(p-1)}(M)$ is the coderivative defined as $\delta\doteq *^{-1}d*$, $*$ being the Hodge-dual. A remarkable property of these operators is that they can be combined together in the Laplace-de Rham operator $\square\doteq d\delta+\delta d$ which coincides with $-\square_g=-g^{\mu\nu}\nabla_\mu\nabla_\nu$, minus the wave operator. Hence, \eqref{eom} can be rewritten as 
$$(\square - d\delta + m^2) A = 0,$$
which, in an arbitrary frame of $(M,g)$, assumes the more renown form:
\beq\label{eomlocal}
(\square_g-m^2) A_\mu + \nabla_\mu\nabla^\nu A_\nu - R_\mu^\nu A_\nu=0,
\eeq
$R_{\mu\nu}$ being the Ricci tensor associated to $g$. 

At first glance this equation does not look hyperbolic, but if one applies the codifferential to \eqref{eom}, one obtains:
$$\delta(\delta d+m^2)A=m^2\delta A = 0.$$
Whenever $m\neq 0$, this entails the Lorentz gauge condition since $\delta A=0$ implies in a local coordinate $\nabla^\mu A_\mu=0$. Hence, every solution of \eqref{eom} is also coclosed, which guarantees us that the dynamics can be equivalently described by a second order hyperbolic partial differential equation:
\beq\label{eomgauge}
(\square+m^2)A=0,\qquad \delta A=0.
\eeq

The set of solutions of the first equation in \eqref{eomgauge} can be generated with the help of the so-called advanced $(^- )$ and retarded $(^+ )$ fundamental solutions $E^\pm_m:\Omega^1_0(M)\to\Omega^1(M)$ where $\Omega^1_0(M)$ represents the set of smooth and compactly supported one-forms on $M$. The operators $E^\pm_m$ are left and right inverses of $\square+m^2$ on the appropriate functional spaces and they yield the standard support properties, {\it i.e.}, $\textrm{supp}(E_m^\pm(f))\subseteq J^\pm(\textrm{supp}(f))$ for all $f\in\Omega^1_0(M)$. Furthermore since $[\square,\delta]=[\square,d]=0$, $E^\pm_m$ intertwine the action of both $\delta$ and $d$. This allows to write the space of solutions for \eqref{eomgauge} as
$$\mathcal{S}_m(M)=\left\{A_f\in\Omega^1(M)\;|\;\exists f\in\Omega^1_0(M)\;\textrm{and}\; A_f=\Delta(f)\doteq \Delta^+(f)-\Delta^-(f)\right\},$$
where $\Delta^\pm\doteq E^\pm_m\left(\mathbb{I}+m^{-2}d\delta\right)$, $\mathbb{I}$ being the identity operator. See also \cite{Fewster} although a different signature for the metric is employed. Notice that the Lorenz gauge condition is automatically implemented since, for all $f\in\Omega^1_0(M)$, the following chain of identities holds:
$$\delta\Delta(f)=\delta\left(E_m(\mathbb{I}+m^{-2}d\delta)f\right)=E_m\left((\delta+m^{-2}\delta d\delta)f\right)=
m^{-2} E_m\left((m^2+\square)\delta f\right)=0.$$
Notice that, exactly as for real scalar fields, it is possible to single out the kernel of the causal propagator -- see also lemma A.2 in \cite{Fewster} -- introducing the following set of equivalence classes: 
$$\left[\Omega^1_0(M)\right]_m=\{[f]\;|\;f\in\Omega^1_0(M),\;{\rm and}\; f\sim f'\Longleftrightarrow\exists\tilde f\in\Omega^1_0(M)\;|\;f-f'=(\delta d+m^2)\tilde f\}.$$
The space of solutions can be rewritten as $\mathcal{S}_m(M)\equiv\Delta\left(\left[\Omega^1_0(M)\right]_m\right)$ and, on account of the results of section 4.3 in \cite{BGP}, pg. 129 in particular, it turns out that $\mathcal{S}_m(M)$ is a weakly non-degenerate symplectic vector space if endowed with the following antisymmetric bilinear form 
\beq\label{symplmass}
\sigma_m(A_f,A_h)\doteq\int\limits_M \Delta(f)\wedge *h.\quad\forall A_f,A_h\in\mathcal{S}_m(M)
\eeq
Notice that both $f$ and $h$ are meant as arbitrary representatives of the equivalence classes in $[\Omega^1_0(M)]_m$ generating $A_f$ and $A_h$ respectively; the symplectic form is manifestly independent from such a choice.

With all these data it is possible to apply a standard construction to associate an algebra of observables to a massive spin $1$ field:
{\definition\label{Weylmassive} We call {\bf Weyl algebra of a Proca field} on a four dimensional globally hyperbolic spacetime $(M,g)$ the unique (up to $*$-isometries) $C^{ *}$-algebra $\mathcal{W}_m(M,g)$ associated to $(\mathcal{S}_m(M),\sigma_m)$ and generated by the elements $W(A_f)$ for all $A_f\in\mathcal{S}_m(M)$ together with the defining relations:
$$W(A_f)^*=W(-A_f),\qquad W(A_f)W(A_h)=e^{\frac{i}{2}\sigma_m(A_f,A_h)}W(A_f+A_h).$$
Furthermore this algebra satisfies the time-slice axiom, that is, if $\Sigma$ is a Cauchy surface of $(M,g)$ and $\mathcal{O}$ an open globally hyperbolic subset of $(M,g)$ containing $\Sigma$, it holds
$$\mathcal{W}_m(M,g)=\mathcal{W}_m(\mathcal{O},g|_{\mathcal{O}}).$$}
The last statement of this definition is a direct consequence of lemma A.3 in \cite{Fewster}. 

\vskip .4cm

\noindent{\it The dynamics and the Weyl algebra of massless spin $1$ fields:} The scenario, in which \eqref{eom} is taken with $m=0$, is more involved and thus it requires a careful analysis on its own. Already the equation of motion could be seen as originating from two different paths:
\begin{itemize}
\item The first calls for considering the curved spacetime analogue of Maxwell equations, that is a two-form, called field strength, $F\in\Omega^2(M)$ which fulfils $dF=0$ and $\delta F=0$. Unless the second cohomology group $H^2(M)$ is trivial, we cannot apply Poincar\'e lemma to conclude that there exists a globally defined $A\in\Omega^1(M)$ such that $F=dA$. This obstruction entails that not all field strength tensors which are dynamically admissible descend from a global vector potential; yet those which can be constructed in this way originate from vector potentials $A\in\Omega^1(M)$ obeying the equation of motion $\delta dA=0$. For a recent analysis focused on the field strength tensor, refer to \cite{Lang}.
\item The second calls for considering the vector potentials $A\in\Omega^1(M)$ as the building block of the theory. Their dynamics are ruled by the action $S[A]=\frac{1}{4}\int\limits_M dA\wedge *dA$, where $*$ is the Hodge dual. The associated Euler-Lagrange equation is $\delta dA=0$ and accordingly we define:
$$\mathcal{M}\doteq\left\{[A],A\in\Omega^1(M),\;\delta dA=0\;\text{and}\;A\sim A'\Longleftrightarrow\exists\Lambda\in\Omega^1(M),\;d\Lambda=0\;\text{and}\;A-A'=\Lambda\right\}.$$ 
\end{itemize}

In this paper we shall focus on the second scenario even though all vector potentials dynamically allowed might not exhaust the possible set of field strengths as if we were considering $F$ as the main object. Nonetheless, even in this picture $dA$ still remains the basic observable and this entails the existence of a gauge freedom, namely two vector potentials which differ by a smooth and closed $1$-form are indistinguishable. This justifies the definition of $\mathcal{M}$.

Such freedom plays a further key role also in the study of the classical dynamics of a vector potential: Contrary to the massive case, $\delta A=0$ is not identically satisfied by all solution of \eqref{eom} with $m=0$. Yet the following lemma avoids any possible complication:
{\lemma\label{gaugesol} Every solution of \eqref{eom} is gauge equivalent to one of 
\beq\label{eom0}
\square A=0, \qquad\delta A=0,
\eeq
where $A\in\Omega^1(M)$.} 
\begin{proof}
By direct inspection every solution of \eqref{eom0} solves $\delta dA=0$. Hence, let us consider any $A$ solving the latter equation and we look for $\Lambda\in\Omega^1(M)$ such that $d\Lambda=0$ and $\square(A+\Lambda)=0$, $\delta(A+\Lambda)=0$. In particular let us choose $\chi\in C^\infty(M)$ and let us fix $\Lambda=-d\chi$; in order to guarantee the coclosedness of $A'\doteq A-d\chi$, it must hold $\delta A = \delta d\chi= \square\chi$. This constraint on $\chi$ entails that $A'$ solves automatically the wave equation since $\square A' =\square A - \square d\chi = \square A -d\square\chi = \square A - d\delta A=\delta dA=0$. To conclude we notice that the existence of at least one function $\chi$ satisfying the above requisites descends from corollary $5$ in chapter 3 of \cite{BF}; it guarantees the existence of a smooth solution of a scalar equation of the form $\square\chi=f$ with $f\in C^\infty(M)$ provided that smooth initial data are assigned. 
\end{proof}
\noindent Yet since the actual physical system is represented by the elements of $\mathcal{M}$, we cannot simply focus on the solutions of \eqref{eom}, but we have to identify those vector potential solving \eqref{eom0} and lying in the same equivalence class of $\mathcal{M}$:

{\lemma\label{bije} Let $\mathcal{L}$ be the set of equivalence classes of vector potentials solving \eqref{eom0} where $A'\sim A$ if and only if there exists $\Lambda\in\Omega^1(M)$ such that $d\Lambda=0$, $\delta\Lambda=0$ and $A'=A+\Lambda$. Then there exists a bijection $\varphi:\mathcal{M}\to\mathcal{L}$ such that $\varphi([A])=[A']$ where $[A']$ is the equivalence class in $\mathcal{L}$ generated by the $1$-form $A'=A+d\chi$ where $\chi\in C^\infty(M)$ and $\square\chi=\delta A$.}

\begin{proof}
The map $\varphi$ is well-defined since lemma \ref{gaugesol} guarantees that $A'$ solves \eqref{eom0} and thus $[A']\in\mathcal{L}$. Furthermore $[A']$ does not depend on $\chi$. As a matter of fact we have the freedom to choose $\chi,\chi'\in C^\infty(M)$ such that $\square\chi=\square\chi'=\delta A$, but $d\chi\neq d\chi'$. Yet $A'=A-d\chi=A-d\chi'-d(\chi-\chi')$ and, if we consider $\Lambda\doteq d(\chi'-\chi)$ it is immediate that $d\Lambda=0$ and $\delta\Lambda=\square(\chi'-\chi)$ which vanishes. 

We show that $\varphi$ is a bijection. 
Since the map is linear, injectivity of $\varphi$ is guaranteed if we show that only $[0]_\mathcal{M}\in Ker(\varphi)$. Suppose this is not true, than there would exist $[A]\in\mathcal{M}$ such that $\varphi([A])=[0]_{\mathcal{L}}$. This is not possible since, per definition of $\varphi$, there exists $\chi\in C^\infty(M)$ such that $A+d\chi=0$ and $\square\chi=\delta A$. Yet, on account of the definition of $\mathcal{M}$, this entails that $A\in [0]_{\mathcal{M}}$.  Surjectivity instead can be easily proven noticing that every $[A]\in\mathcal{L}$ is generated by $A$, solution of \eqref{eom0} and thus also of \eqref{eom}. Hence $A$ generates also an equivalence class in $\mathcal{M}$ and any other representative of $[A]\in\mathcal{L}$ would fall in the same equivalence class since two elements differ by a smooth $1$-form which is both closed and co-closed. To conclude, a direct computation shows that $\varphi\circ\varphi^{-1}=id:\mathcal{L}\to\mathcal{L}$ while $\varphi^{-1}\circ\varphi=id:\mathcal{M}\to\mathcal{M}$. 
\end{proof}

\vskip .2cm

As a by-product of this lemma we can focus only on the solutions of \eqref{eom0} and, as usual, we are interested in those generated by compactly supported initial data. In order to account automatically for the Lorenz condition, we use a procedure first implemented in \cite{Dimock}, see proposition 4 in particular: We define as space of test functions:
$$\Omega^1_{0,\delta}(M)\doteq\{f\in \Omega^1_0(M)\;\textrm{such that}\; \delta f=0\},$$
from which we construct
$$\mathcal{S}(M)=\left\{A\in\Omega^1(M)\;|\;\exists f\in \Omega^1_{0,\delta}(M)\;|\; A=E(f)\right\},$$
where $E$ is the causal propagator associated to the $\square$-operator. Yet, on account of the definition of $\mathcal{L}$, it is still possible to assign two distinct initial data yielding two $1$-forms which are in the same equivalence class and we want to single out such possibility.

{\definition\label{equivrel} We call $\left[\Omega^1_{0,\delta}(M)\right]$ the set of equivalence classes $[f]$ constructed out of the following equivalence relation:
\begin{itemize}
\item $f\sim f'$ if $f,f'\in\Omega^1_{0,\delta}$ and if there exists $\beta\in\Omega^2_0(M)$ such that $f-f'=\delta\beta$ and $d\beta=0$.
\end{itemize}
Then we define the space of Lorentz solutions with compactly supported initial data $\mathcal{L}_0(M)\doteq E\left[\Omega^1_{0,\delta}(M)\right]$, where $E$ is the causal propagator of $\square$ in $(M,g)$. We also introduce $\mathcal{M}_0(M)\doteq\varphi^{-1}\left(\mathcal{L}_0(M)\right)$, $\varphi$ being the map defined in lemma \ref{bije}.}\\

Notice that the equivalence relation between two test functions can be justified as follows: Let $f,f'$ be any two representatives of $[f]\in\left[\Omega^1_{0,\delta}(M)\right]$. Then $E(f)-E(f')=E(f-f')=E(\tilde f)$ where $\tilde f\in\Omega^1_{0,\delta}(M)$; hence $\delta E(\tilde f)=0$. Furthermore, since $f$ and $f'$ should generate two solutions in the same equivalence class of $\mathcal{L}$, it must hold $dE(\tilde f)=E(d\tilde f)=0$, which entails that there exists $\beta\in\Omega^2_0(M)$ such that $d\tilde f=\square\beta$. Yet $\square\tilde f=d\delta f+\delta d\tilde f=\delta\square\beta=\square\delta\beta$, which, due to the compactness of all forms, yields $\tilde f=\delta\beta$. Furthermore, since $d\tilde f=\square\beta$, it must also hold that $\delta d\beta=0$, which in turn implies the identity $\square d\beta=0$, which is equivalent to $d\beta=0$, $\beta$ being compactly supported. 

In order to discuss the quantization of the vector potential, we can proceed constructing an associated Weyl algebra. To this avail we need to identify a suitable symplectic form on $\mathcal{M}$ or rather, in view of the preceding discussion, on $\mathcal{M}_0$. The auxiliary space of Lorentz solutions is rather handy since

{\proposition The set $\mathcal{L}_0(M)$ is a symplectic vector space if $H^1(M)$ is trivial and if endowed with the following weakly non-degenerate symplectic form:
\beq\label{sympl}
\sigma([A_f],[A_h])=\int\limits_M E(f)\wedge *h,
\eeq
where $f,h$ are any representative of $[f],[h]\in\left[\Omega^1_0(M)\right]$ while $*$ stands for the Hodge-operator. This automatically induces a weakly non-degenerate symplectic form $\sigma_{\mathcal{M}_0}:\mathcal{M}_0\times\mathcal{M}_0\to\mathbb{R}$ defined as the pull-back
$$\sigma_{\mathcal{M}_0}([A],[A'])\doteq\varphi^*\sigma([A],[A'])\doteq\sigma(\varphi[A],\varphi[A']).\qquad\forall [A],[A']\in\mathcal{M}_0$$
}
\begin{proof}
As a starting point let us notice that the right hand side of \eqref{sympl} contains specific representatives of the equivalence classes appearing on the left hand side. Hence, in order for \eqref{sympl} to be well-defined, one must prove that the right hand side is independent from the choice of the representative. Since $\sigma$ is bilinear and antisymmetric, we can just consider one of the arguments and, thus, suppose that $f,f'\in[f]\in\mathcal{L}_0$. Hence $f-f'=\tilde f\in\Omega^1_{0,\delta}(M)$ with $\tilde f=\delta\beta$, $\beta$ being a closed $2$-form of compact support. Since $dE(\tilde f)=0$ and since the first cohomology group of $M$ is trivial, we know that there exists $\lambda\in C^\infty(M)$ such that $E(\tilde f)=d\lambda$. Therefore
$$\int\limits_M E(\tilde f)\wedge *h=\int\limits_M d\lambda\wedge *h=
\int\limits_M \lambda\wedge *\delta h=0,$$
where in the last identity we used the fact that the intersection of the support of $\lambda$ and $h$ is compact. Notice that, without the topological assumption on the cohomological structure of $M$, there is no apparent way to claim that $\int\limits_M E(\tilde f)\wedge *h$ vanishes even knowing that $\tilde f$ is both closed and coclosed. Since $\varphi$ is a bijection as per proposition \ref{bije} and since $\mathcal{M}_0$ is per definition the pre-image of $\mathcal{L}_0$, we only need to prove that $(\mathcal{L}_0,\sigma)$ is a weakly non-degenerate symplectic vector space. This is tantamount to show that, if $\sigma([A_f],[A_h])=0$ for all $[A_h]\in\mathcal{L}_0$ then $[A_f]$ must be $[0]$. Let us suppose that this is not true and let us assume that $h=\delta\widetilde\lambda$ with $\widetilde\lambda\in \Omega^2_0(M)$. This entails that 
$$\int\limits_M E(f)\wedge *\delta\lambda=\int\limits_M dE(f)\wedge *\widetilde\lambda=0.$$
Since both $d E(f)$ and $*\widetilde\lambda$ are two-forms, the non-degenerateness of the pairing between $2$-forms and the arbitrariness of $\lambda$ yield that $dE(f)=0$. Since per hypothesis also $\delta E(f)$ vanishes, it means that $E(f)$ falls in the same equivalence class as $0$ in $\mathcal{L}$. Since $f$ is also compactly supported, the desired conclusion follows.
\end{proof}

\noindent We are now in position to define the algebra of observables for a massless spin $1$ field:

{\definition\label{Weylmassless} We call {\bf Weyl algebra} of the spin $1$ massless field on $(M,g)$ with $H^1(M)=\{0\}$, the unique (up to *-isometries) C$^{\,*}$-algebra $\mathcal{W}_0(M,g)$ associated to the symplectic space $(\mathcal{M}_0,\sigma_{\mathcal{M}_0})$ which is generated by the abstract elements $W([A])$ with $[A]\in\mathcal{M}_0$ and such that 
$$W([A])^*=W(-[A]),\qquad W([A])W([A'])=e^{\frac{i}{2}\sigma_{\mathcal{M}_0}([A],[A'])}W([A+A']).$$
}

\noindent Whenever the first cohomology group of $(M,g)$ is not trivial, even though the straightforward construction of a Weyl C$^*$-algebra fails, one could try to resort to an alternative procedure which stems from the existence of a cover of any globally hyperbolic spacetime with contractible globally hyperbolic subsets. Since on each such subset all topological obstructions would be absent and thus each of them possesses its own Weyl algebra, one could try to associate to $(M,g)$ the so-called universal algebra using the scheme outlined in \cite{Fredenhagen}. Yet, in order to prove the existence of such an algebra one would have to verify that the conditions of proposition B.0.6 in \cite{Brunetti} are met; these appear to be rather tricky to prove and we postpone the study of this problem to a later work. 

\vskip .2cm

\noindent Although already proven in proposition A.3 of \cite{Fewster}, we recast in our language the following result:

{\lemma The Weyl algebra $\mathcal{W}_0(M,g)$ satisfies the {\bf time-slice axiom}, that is, if $\Sigma$ is a Cauchy surface of $(M,g)$ and $\mathcal{O}$ an open globally hyperbolic subset of $M$ containing $\Sigma$, it holds that $\mathcal{W}_0(M,g)=\mathcal{W}_0(\mathcal{O},g|_{\mathcal{O}})$.}
\begin{proof}
Let us recall that every generator $W([A_f])$ of $\mathcal{W}(M,g)$ can be constructed unambiguously from $\left[\Omega^1_{0,\delta}(M,g)\right]$. Hence we can equivalently consider the C$^*$-algebra generated by $V([f])$ with $[f]\in\left[\Omega^1_{0,\delta}(M,g)\right]$ and subject to the defining relations:
$$V([f])^*=V(-[f]),\qquad V([f])V([f'])=e^{\frac{i}{2}E([f],[f'])}V([f+f']),$$
where $E([f],[f'])\doteq\int\limits_M E(f)\wedge *f'$, $f,f'$ being arbitrary representatives in the respective equivalence classes. Let us now consider an arbitrary globally hyperbolic open subset $\mathcal{O}$ of $M$ encompassing a Cauchy surface $\Sigma$ and let us choose two other Cauchy surfaces $\Sigma^\pm$, lying in the future and in the past of $\Sigma$ respectively. Let us also fix two functions $\chi^\pm\in C^\infty(M)$ such that $\chi^++\chi^-=1$ and $\textrm{supp}(\chi^+)\in J^-(\Sigma^+)$ while $\textrm{supp}(\chi^-)\in J^+(\Sigma^-)$. Hence consider now any $[f]\in\left[\Omega^1_{0,\delta}(M,g)\right]$ and suppose there exists a representative $f\in\Omega^1_{0,\delta}$ supported in the causal future of $\Sigma^+$. Then the following identity holds: 
$$f=\delta d(E^-(f)-\chi^+E(f))+\tilde f.$$
Hence, $\tilde f=f-\delta d(E^-(f)-\chi^+E(f))=\delta d(\chi^+E(f))$, where we exploited that $\delta f$ vanishes. Thanks to the support properties of $\chi^\pm$, it turns out that $\tilde f\in\Omega^1_0(\mathcal{O})\subset\Omega^1_0(M)$. Furthermore, it also holds that $E(f)=E(\tilde f)$ and, since $\delta f=0$, also $\delta \tilde f=0$. To conclude we need to show that $\tilde f$ falls in the same equivalence class as $f$ in $(M,g)$ as per definition \ref{equivrel}: This is automatic since, per construction, $f-\tilde f=\delta\beta$ where $\beta=d(\chi^+E(f))$ and $d\beta=0$.
\end{proof}

\subsection{Spin $\frac{1}{2}$ and spin $1$ fields in the language of general local covariance}

We remind the reader that our ultimate goal is to prove the existence of states for higher spin free fields enjoying at least locally the Reeh-Schlieder property. Particularly we want to follow the approach of \cite{Sanders2} which has the advantage of being rather general since it deals with arbitrary locally covariant field theories. First introduced in \cite{BFV}, the so-called principle of general local covariance was formulated leading to the realization of a quantum field
theory as a covariant functor between the category of globally hyperbolic (four-dimensional) Lorentzian manifolds with isometric embeddings as morphisms and the category of C$^*$-algebras with invertible endomorphisms as morphisms. This also entails a new interpretation of local fields as natural transformations from compactly supported smooth functions to suitable operators. In the same paper it was proven that the Klein-Gordon field fits perfectly in this scheme whereas Dirac fields were dealt with in \cite{Sanders3}, although a preliminary analysis was already present in \cite{Verch}. Therefore, the goal of this subsection will be twofold: On the one hand we will shorty recapitulate some of the results of this last cited paper concerning spin $\frac{1}{2}$ fields, while, on the other hand, we will prove that also the vector potential and the Proca field can be described as genuine local covariant quantum field theories. Since, as already mentioned, the natural language of this framework is that of categories, as a first step we will introduce those which will be employed in the forthcoming analysis. Since the composition map between morphism and the existence of an identity map are straightforwardly defined in all the cases we shall consider, we will omit them.

{\definition We call:
\begin{itemize}
\item $\mathfrak{GlobHyp}$: the category whose objects are $(M,g)$, that is four dimensional oriented and time oriented globally hyperbolic spacetimes endowed with a smooth metric of signature $(-,+,+,+)$. A morphism between two objects $(M,g)$ and $(M',g')$ is a smooth embedding $\mu:M\to M'$ such that $\mu(M)$ is causally convex\footnote{We recall that an open subset $\mathcal{O}$ of a globally hyperbolic spacetime is called {\em causally convex} if $\forall x,y\in\mathcal{O}$ all causal curves connecting $x$ to $y$ lie entirely inside $\mathcal{O}$.} and $\mu^*g'=g$ on $\mu(M)$.
\item $\mathfrak{GlobHyp}_1$: the subcategory of $\mathfrak{GlobHyp}$ whose objects are $(M,g)\in{\rm Obj}(\mathfrak{GlobHyp})$ and $H^1(M)=\{0\}$. A morphism between two objects $(M,g)$ and $(M',g')$ is a smooth embedding\footnote{Notice that $\mu$ is a diffeomorphism between $M$ and $\mu(M)$ and thus corollary 11.3 of \cite{Lee} entails that the cohomology groups of $M$ and $\mu(M)$ are isomorphic.} $\mu:M\to M'$ such that $\mu(M)$ is causally convex and $\mu^*g'=g$ on $\mu(M)$.
\item $\mathfrak{Bund}$: the category whose objects are smooth fibre bundles $\pi:P\to M$ where $(M,g)$ is an object of $\mathfrak{GlobHyp}$. Morphisms are smooth maps $\nu:P_1\to P_2$ such that they restrict to isomorphisms of the fibres and they cover a morphism $\mu:M_1\to M_2$ in $\mathfrak{GlobHyp}$, that is $\pi_2\circ\nu=\mu\circ\pi_1$. 
\item $\mathfrak{SSpac}$: the subcategory of $\mathfrak{Bund}$ whose objects are the quadruples $(M,g,SM,\pi_S)$ where $(M,g)$ is an object of $\mathfrak{GlobHyp}$ whereas $(SM,\pi_S)$ is a spin bundle over $(M,g)$ as per definition \ref{spinstr}. Morphisms are maps $\chi:(M_1,g_1,SM_1,\pi_{S,1})\to (M_2,g_2,SM_2,\pi_{S, 2})$ covering morphisms $\mu:(M_1,g_1)\to (M_2,g_2)$ in $\mathfrak{GlobHyp}$ so that $\chi\circ R_{S,1}=R_{S,2}\circ\chi$ and $\pi_{S,2}\circ\chi=\mu_*\circ\pi_{S,1}$ where $\mu_*$ is the push-forward induced by $\mu$.
\item $\mathfrak{Alg}$: the category whose objects are unital C$^{\,*}$-algebras whereas morphisms are injective unit-preserving $*$-homomorphisms.
\end{itemize}}


We shall now use these ingredients first discussing the Proca field and the vector potential, which are still treated separately due to some subtleties, and later recollecting the results of \cite{Sanders2} on Dirac fields. Hence, on account of definition 2.1 in \cite{BFV}:

{\proposition\label{LCQFTm} The Proca field is a locally covariant quantum field theory, that is there exists a covariant functor ${\rm W}_{A,m}:\mathfrak{GlobHyp}\to\mathfrak{Alg}$ which assigns to every object $(M,g)\in\mathfrak{GlobHyp}$ the $C^*$-algebra $\mathcal{W}_m(M,g)$ of definition \ref{Weylmassive} with the induced action on the morphisms. In diagrammatic form:
\begin{equation*}
\begin{CD}
 (M,g)@>{\mu}>> (M',g') \\
 @V{{\rm W}_{A,m}}VV        @VV{{\rm W}_{A,m}}V\\
\mathcal{W}_m(M,g) @>{\alpha_\mu}>> \mathcal{W}_m(M',g')
\end{CD}
\end{equation*}
where $\alpha_\mu$ is the unit-preserving $*$-homomorphism defined by its action on the generators as $\alpha_\mu\left(W(A_f)\right)\doteq W(A_{\tilde{f}})$ where $\tilde{f}\doteq f\circ\mu$ for all $A_f\in\mathcal{S}_m(M)$.
Furthermore the locally covariant quantum field theory of a Proca field fulfils the {\bf time slice axiom} and it is {\bf causal}, that is for every two morphisms $\mu_j:(M_j,g_j)\to(M,g)$, $j=1,2$ between objects in $\mathfrak{GlobHyp}$ so that $\mu_1(M_1,g_1)$ is causally separated from $\mu_2(M_2,g_2)$, it holds 
$$\left[\alpha_{\mu_1}\left(\mathcal{W}_m(M_1,g_1)\right),\alpha_{\mu_2}\left(\mathcal{W}_m(M_2,g_2)\right)\right]=0.$$
}
\begin{proof}
On account of the analysis of the previous section, the proof is roughly identical to the one for the Klein-Gordon field given in \cite{BFV}, which in turn relies on the analysis of \cite{Dimock3}. As we discussed in definition \ref{Weylmassive} and in the preceding analysis, to every $(M,g)\in\textrm{Obj}(\mathfrak{GlobHyp})$, one can associate both $(\mathcal{S}_m(M),\sigma_m)$, the symplectic vector space built out of the solutions of the Proca equation and $\mathcal{W}_m(M,g)$, a unique (up to $*$-isomorphisms) C$^*$-algebra. Let us consider any morphism $\mu$ between two objects, $(M,g)$ and $(M',g')$ and let us consider $(\mu(M),g'|_{\mu(M)})$ as a globally hyperbolic spacetime on its own. It is easy to see that $\mathcal{S}_m(\mu(M))=\mu_*\mathcal{S}_m(M)$, where $\mu_*$ is the pull-back on forms induced by the action of $\mu^{-1}$ which is well-defined, $\mu$ being a diffeomorphism between $M$ and $\mu(M)$. As a matter of fact, if we consider any solution $A$ of the Proca equation, the following identity holds: 
$$\mu_*[(\delta d-m^2)A]=(\delta d- m^2)\mu_*(A)=0,$$
where we employed that the pull-back induced by a smooth isometric embedding commutes with $d$ (see Lemma 9.14 in \cite{Lee}) and it intertwines the Hodge star $*$ of the source and of the target manifold. This identity together with the relations $\mu_*\circ\mu^*=id_{\mu(M)}$ and $\mu^*\circ\mu_*=id_M$ suffices to prove the above statement. Hence, if we consider the propagator $\Delta=E_m(\mathbb{I}+m^{-2}d\delta)$ for the Proca equation in $(M,g)$ and $\widetilde\Delta$ for the counterpart in $(\mu(M),g'|_{\mu(M)})$, it holds similarly to the scenario of a real scalar field that $\widetilde\Delta=\mu_*\circ\Delta\circ\mu^*$. Consequently, on account of definition \eqref{symplmass}, for all $f,h\in\Omega^1_0(M)$
$$\int_M \Delta(f)\wedge *h=\int_{\mu(M)}\widetilde\Delta(\tilde f)\wedge *\tilde h,$$
where $\tilde f$ and $\tilde h$ are equal to $\mu_*(f)$ and $\mu_*(h)$ respectively. In other words $\mu$ is a symplectomorphism. According to a standard theorem for C$^*$-algebras, this entails the existence of a C$^*$-isomorphism $\tilde\alpha_\mu$ between the Weyl algebras $\mathcal{W}_m(M,g)$ and $\mathcal{W}_m(\mu(M),g'|_{\mu(M)})$ whose action on each generator $W(A_f)$, $A_f\in\mathcal{S}_m(M)$, is unambiguously fixed as $\tilde\alpha_\mu(W(A_f))\doteq W(\mu_*(A_f))$. Furthermore since $\mu(M)$ is a globally hyperbolic open subset of $M'$, the uniqueness of the causal propagator for the Proca equation entails that $\widetilde\Delta\equiv\chi(\mu(M))\Delta'$, $\chi(\mu(M))$ and $\Delta'$ being the characteristic function of $\mu(M)\subseteq M'$ and the propagator for $\delta d-m^2$ in $(M',g')$ respectively. This relation yields the existence of a natural immersion $\iota:\mathcal{S}_m(\mu(M))\to\mathcal{S}_m(M')$ which associates to each $A_f\in\mathcal{S}_m(\mu(M))$ with $f\in\Omega^1_0(\mu(M))$ the $1$-form $\Delta'(f)\in\mathcal{S}_m(M')$. A direct inspection of \eqref{symplmass} shows that $\iota$ is a symplectomorphism and thus it induces a $*$-isomorphism $\alpha_\iota:\mathcal{W}_m(\mu(M),g'|_{\mu(M)})\to\mathcal{W}_m(M',g')$ characterized by its action on the generators as $\alpha_\iota(W(A_f))\doteq W(\iota(A_f))$. Hence we have constructed a C$^*$-isomorphism $\alpha_\mu\doteq\alpha_\iota\circ\tilde\alpha_\mu:\mathcal{W}_m(M,g)\to\mathcal{W}_m(M',g')$ and it automatically satisfies the covariance properties required in definition 2.1 of \cite{BFV}, namely 
$$\alpha_{\mu'}\circ\alpha_\mu=\alpha_{\mu'\circ\mu},\qquad\alpha_{id_M}=id_{\mathcal{W}_m(M,g)},$$
where $\mu$ is a morphism between $(M,g)$ and $(M',g')$ in $\mathfrak{GlobHyp}$ while $\mu'$ is a morphism between $(M',g')$ and $(M'',g'')$ in $\mathfrak{GlobHyp}$. This suffices to prove that ${\rm W}_{A,m}$ is indeed a covariant functor defining a locally covariant quantum field theory. As already mentioned, the time-slice axiom holds true as proven by Fewster and Pfenning in lemma A.3 of \cite{Fewster} whereas the property of being causal is a by-product of the composition rule of the Weyl algebra in definition \ref{Weylmassive}. Such rule depends on the symplectic form $\sigma_m$ which in turn is constructed out of the causal propagator $E_m$ for $\square+m^2$. Hence, if we consider two objects $(M_1,g_1)$ and $(M_2,g_2)$ of $\mathfrak{GlobHyp}$ isometrically embedded in a third one, $(M,g)$ and such that $\mu_1(M_1)$ is causally separated with $\mu_2(M_2)$, it holds that, for every $W(A_f)\in\alpha_{\mu_1}(\mathcal{W}_m(M_1,g_1))$ and $W(A_h)\in\alpha_{\mu_2}(\mathcal{W}_m(M_2,g_2))$, $W(A_f)W(A_h)=W(A_f+A_h)=W(A_h)W(A_f)$ since $\sigma_m(A_f,A_h)$ vanishes. This is due to the support properties of the causal propagator $E_m$ and thus also of $\Delta$.
\end{proof}

\noindent We can now focus on the massless case:
{\proposition\label{LCQFT0} The vector potential can be described in terms of a locally covariant quantum field theory, that is there exists a covariant functor ${\rm W}_{A,0}:\mathfrak{GlobHyp}_1\to\mathfrak{Alg}$ which assigns to every object $(M,g)\in\mathfrak{GlobHyp}_1$ the $C^*$-algebra $\mathcal{W}_0(M,g)$ of definition \ref{Weylmassless} with the induced action on the morphisms. In diagrammatic form:
\begin{equation*}
\begin{CD}
 (M,g)@>{\mu}>> (M',g') \\
 @V{{\rm W}_{A,0}}VV        @VV{{\rm W}_{A,0}}V\\
\mathcal{W}_0(M,g) @>{\alpha_\mu}>> \mathcal{W}_0(M',g')
\end{CD}
\end{equation*}
where $\alpha_\mu$ is the unit-preserving $*$-homomorphism defined by its action on the generators as $\alpha_\mu\left(W([A_f])\right)\doteq W([A_{\tilde{f}}])$ where $[\tilde{f}]\doteq [f\circ\mu]$ for all $f\in\left[\Omega^1_{0,\delta}(M,g)\right]$
Furthermore this locally covariant quantum field theory fulfils the {\bf time slice axiom} and it is {\bf causal}.
}

\begin{proof}
Let us start from an arbitrary $(M,g)\in\textrm{Obj}(\mathfrak{GlobHyp}_1)$ and let us consider any smooth isometric embedding $\mu:(M,g)\to(M',g')$. On the image $\mu(M)$, the map $\mu$ is a diffeomorphism and thus, for every $A\in\Omega^1(M)$ solving \eqref{eom0} it is meaningful to consider the pull-back under the action of $\mu^{-1}$ which we indicate as $\mu_*(A)$. As already discussed in the proof of proposition \ref{LCQFTm}, both the exterior derivative and $\delta$ commute with any isometry; thus it holds both that $\square\mu_*(A)=\mu_*(\square(A))=0$ and that $\delta\mu_*(A)=\mu_*\delta(A)=0$. Furthermore, if $A$ can be written as $E(f)$ with $f\in\Omega^1_{0,\delta}$, we can proceed as for the Proca equation to conclude that $\mu_*(A)=\mu_*\circ E(f)=\mu_*\circ E\circ\mu^*\circ\mu_*(f)=\widetilde E\tilde f$ is a solution of \eqref{eom0} in $\mu(M)$ generated by the action on $\tilde f\doteq\mu_*(f)\in\Omega^1_0(\mu(M))$ of the causal propagator $\widetilde E\doteq\mu_*\circ E\circ\mu^*$. Furthermore, on account of definition \ref{equivrel} and of $[\delta,\mu_*]=0$, it holds that $\left[\Omega^1_{0,\delta}(\mu(M))\right]=\mu_*\left[\Omega^1_{0,\delta}(M)\right]$ and thus, $\mathcal{L}_0(\mu(M))=\mu_*\mathcal{L}_0(M)$. For any $[\tilde f],[\tilde h]\in\left[\Omega^1_{0,\delta}(\mu(M))\right]$, \eqref{sympl} reads
\begin{gather*}
\widetilde\sigma\left([A_{\tilde{f}}],[A_{\tilde{h}}]\right)=\int\limits_{\mu(M)}\widetilde E(\tilde f)\wedge *\tilde{h}=
\int\limits_{\mu(M)}\mu_*\circ E(f)\wedge \mu_*(h)=\\
=\int\limits_{\mu(M)}\mu_*\left[E(f)\wedge h\right]=\int_M E(f)\wedge *h=\sigma\left([A_f],[A_h]\right),
\end{gather*}
where $\widetilde\sigma$ stands for the symplectic form of $\mathcal{L}_0(\mu(M))$. Notice that in the various equalities we employed the definitions of $\widetilde E$, $\tilde f$, $\tilde h$ as well as lemma 9.9 in \cite{Lee}. In other words $\mu_*$ is a symplectomorphism between $(\mathcal{L}_0(M),\sigma)$ and $(\mathcal{L}_0(\mu(M)),\widetilde\sigma)$. 

We recall now that $(\mu(M),g'|_{\mu(M)})$ is a globally hyperbolic open subset of $(M',g')$, which describes via the smooth isometric embedding $\iota:\mu(M)\to M'$. Hence, since any $f\in\Omega_{0,\delta}^1(\mu(M))$ also lies in $\Omega^1_{0,\delta}(M')$ every element of the equivalence class $[f]_{\mu(M)}\in\left[\Omega^1_{0,\delta}(\mu(M))\right]$ generated by $f$ lies in the equivalence class $[f]_{M'}\in\left[\Omega^1_{0,\delta}(M')\right]$ generated still by $f$. Notice that the subscripts are introduced in order to avoid confusion in the notation and they will not be present elsewhere in the text. Furthermore the uniqueness of the solution for the wave equation with smooth compactly supported initial data entails that, if we consider $E'$, the causal propagator for \eqref{eom0} in $M'$, it holds that $\widetilde E=\chi[\mu(M)]E'$ where $\chi$ is the characteristic function. Consequently we can claim both that $\iota$ induces an embedding $\tilde\iota:\mathcal{L}_0(\mu(M))\to\mathcal{L}_0(M')$ defined as $\tilde\iota\left(E[f]_{\mu(M)}\right)\doteq E\left([f]_{M'}\right)$ and that $\tilde\iota$ is a symplectomorphism. This can be proven per direct inspection of \eqref{sympl} and thus we omit the explicit computation.

If we now consider the symplectomorphism $\varphi$ from definition \ref{equivrel}, we have proven the existence of a map $\tilde\mu\doteq\varphi^{-1}_{M'}\circ\tilde\iota\circ\mu_*\circ\varphi_M:\mathcal{M}_0(M)\to\mathcal{M'}_0$ which preserves the symplectic form. At a level of Weyl algebras this induces a C$^*$-isomorphism $\alpha_\mu:\mathcal{W}_0(M)\to\mathcal{W}_0(M')$ completely determined by its action on the generators:
$$\alpha_\mu(W([A]))=W(\varphi^{-1}_{M'}\circ\tilde\iota\circ\mu_*\circ\varphi_M([A])),\qquad\forall[A]\in\mathcal{M}_0(M).$$
On account of this last formula , the map $\alpha_\mu$ automatically satisfies the covariance properties required in definition 2.1 of \cite{BFV}, namely 
$$\alpha_{\mu'}\circ\alpha_\mu=\alpha_{\mu'\circ\mu},\qquad\alpha_{id_M}=id_{\mathcal{W}_0(M,g)},$$
where $\mu$ is a morphism between $(M,g)$ and $(M',g')$ in $\mathfrak{GlobHyp}_1$ while $\mu'$ is a morphism between $(M',g')$ and $(M'',g'')$ in $\mathfrak{GlobHyp}_1$. This suffices to prove that ${\rm W}_{A,0}$ is indeed a covariant functor defining a local covariant quantum field theory. As in the massive case, the time-slice axiom is satisfied as proven by Pfenning and Fewster in \cite{Fewster}, whereas causality holds true following exactly the same reasoning as in proposition \ref{LCQFTm} barring minor amendments to account for the different equivalence classes.
\end{proof}

We can now focus on the fields with spin $\frac{1}{2}$. As we mentioned before this case has been treated in full details in \cite{Sanders3} and it is far from our goals to summarize all the results obtained in this cited paper. Bearing in mind the nomenclature of definition \ref{Diracalg} and of the subsequent analysis, here we shall only summarize in our setting the results of theorem 4.7 and proposition 4.8 in \cite{Sanders3}:

{\proposition\label{DCQFT} A Dirac field can be described as a locally covariant quantum field theory, namely as the  ${\rm B}:\mathfrak{SSpac}\to\mathfrak{Alg}$ which assigns to each $(M,g,SM,\pi_S)\in{\rm Obj}(\mathfrak{SSpin})$ $\mathcal{A}(M,g)$, the subalgebra of observables of $\mathfrak{F}(M,g)$. Furthermore this theory is causal and it satisfies the time-slice axiom.}

\section{The Reeh-Schlieder property}
In the previous section we set all the pieces on the chessboard and now it is time to unveil the details of the strategy according to which we shall use them. Hence, as a first step, we shall recollect the notion of ``Reeh-Schlieder'' property.

As customary in the algebraic approach to quantum field theory, a full quantization scheme consists of two ingredients, a unital C$^*$-algebra (a $*$-algebra actually suffices in the most general scenario), here indicated for the sake of simplicity as $\mathcal{W}$, and a state, that is a continuous linear functional $\omega:\mathcal{W}\to\mathbb{C}$ such that 
$$\omega(e)=1,\qquad\omega(a^*a)\geq 0,\quad\forall a\in\mathcal{W},$$
where $e$ is the unit element of the algebra. The Gelfand-Naimark-Segal (GNS) theorem guarantees that it is possible to assign to the pair $(\mathcal{W},\omega)$ a triplet $(\mathcal{H}_\omega,\Pi_\omega,\Omega_\omega)$, where $\mathcal{H}_\omega$ is an Hilbert space on which the algebra $\mathcal{W}$ is represented in terms of bounded linear operators via $\Pi_\omega$. Furthermore, $\Omega_\omega$ is a norm $1$ vector in $\mathcal{H}_\omega$, such that 
$\mathcal{H}_\omega=\overline{\pi_\omega(\mathcal{W})\Omega_\omega}$. The collection of all states admits a description in terms of the following category:
\begin{itemize}
\item $\mathfrak{States}$: The objects are all the subsets of $\mathcal{W}^*$, the dual space of $\mathcal{W}\in\textrm{Obj}(\mathfrak{Alg})$ and as morphisms all maps $\alpha^*:S_1\to S_2$ defined as the pull-back constructed out of the morphism $\alpha:\mathcal{W}_1\to\mathcal{W}_2$ between two objects in $\mathfrak{Alg}$.
\end{itemize}
{\definition\label{statespace} For a given functor ${\rm W}:\mathfrak{GlobHyp}\to\mathfrak{Alg}$ defining a local covariant quantum field theory in the sense of definition 2.1 in \cite{BFV}, a {\bf state space} $S$ is a contravariant functor $S:\mathfrak{GlobHyp}\to\mathfrak{States}$ such that $S(M,g)={\rm W}(M,g)^*$ for all $(M,g)\in{\rm Obj}(\mathfrak{Globhyp})$ whereas, for a given morphism $\mu:(M,g)\to (M',g')$, it holds that $S(\psi)\doteq\alpha^*_\mu$, this being the pull-back induced by $\alpha_\mu:{\rm W}(M,g)\to {\rm W}(M',g')$.

Furthermore, a state $\omega\in S(M,g)$ with $(M,g)\in{\rm Obj}(\mathfrak{GlobHyp})$ has the {\bf Reeh-Schlieder property} for a causally convex region $\mathcal{O}\subseteq M$ if and only if $\overline{\pi_\omega\left({\rm W}(\mathcal{O},g|_{\mathcal{O}})\right)\Omega_\omega}=\mathcal{H}_\omega$.}\\

\noindent Notice that the same definition is valid if we replace $\mathfrak{GlobHyp}$ with $\mathfrak{GlobHyp}_1$ or with $\mathfrak{SSpac}$.

\subsection{The deformation argument}

The next step consists of explaining the procedure leading to prove the existence of a state which satisfies the Reeh-Schlieder property at least for a causally convex region of a globally hyperbolic spacetime. This issue was addressed in \cite{Sanders2} for a generic but abstract local covariant field theory while only the concrete case of a scalar field theory was discussed in detail. The underlying philosophy is the same as the one used in the last decade to prove that there exist Hadamard states for free field theories, namely a deformation argument. First introduced in \cite{FNW}, it calls for mapping via a local isometry a suitable neighbourhood of a Cauchy surface in an arbitrary globally hyperbolic spacetime into an open neighbourhood of a Cauchy surface of a second globally hyperbolic spacetime. The latter is engineered in such a way that its metric becomes isometric to that of an ultrastatic globally hyperbolic spacetime\footnote{A four dimensional globally hyperbolic spacetime is called {\em ultrastatic} if there exists a local chart $(t,x^i)$, $i=1,...,3$ such that the line element reads $ds^2=-dt^2+h_{ij}dx^idx^j$ where $t$ runs over the whole real line while $h$ is a smooth Riemannian metric independent from $t$.} in a causal convex neighbourhood of a third Cauchy surface. The advantage lies in the presence on any ultrastatic spacetime of a complete timelike Killing field which makes possible an explicit construction of states enjoying most of the wanted properties such as the Reeh-Schlieder one or/and the Hadamard condition. Furthermore, the existence of a state with these properties in an ultrastatic manifold suffices to guarantee the existence of a second state in the first manifold which preserves at least locally the very same properties. We review now this procedure more in detail although all the statements we shall write are proven in section 3 of \cite{Sanders2}. Notice that, as in the previous section, we can replace $\mathfrak{GlobHyp}$ with $\mathfrak{GlobHyp}_1$ without problems. The same conclusion holds true in the case of spin $\frac{1}{2}$ fields, where one should work with $\mathfrak{SSpac}$. To this avail we recall that to each object $(M,g,SM,\pi_S)\in\mathfrak{SSpac}$ it is associated a unique element $(M,g)\in\textrm{Obj}(\mathfrak{GlobHyp})$ and that each morphism in $\mathfrak{SSpac}$ is the covering of one in $\mathfrak{GlobHyp}$. One might think that potential subtleties might arise when there does not exists a unique spin structure associated to a four dimensional oriented and time oriented globally hyperbolic spacetime $(M,g)$. Yet this is ultimately not a problem since the mentioned non uniqueness is ruled by the topology of the underlying background. Hence, since the deformation argument focuses only on the geometry leaving the topology untouched, we are free to associate to all the spacetimes involved the same spin structure and thus all the results we shall derive can be used also for Dirac fields once a spin structure has been fixed.

\vskip .2cm

\noindent Hence, the geometric side of the deformation argument is the following:
{\proposition\label{def1} Let $(M,g), (M',g')\in{\rm Obj}(\mathfrak{GlobHyp})$ be chosen so that the respective Cauchy surfaces $\Sigma\hookrightarrow M$, $\Sigma'\hookrightarrow M'$ are diffeomorphic. Let $\mathcal{O}'\subset M'$ be a bounded causally convex (bcc) region with non-empty causal complement\footnote{We recall that for any open set $K$ of a Lorentzian manifold $(M,g)$, the causal complement is $K^\perp\doteq M\setminus\overline{\left(J^+(K)\cup J^-(K)\right)}$.}. Then there exist $(\tilde M,\tilde g)\in{\rm Obj}(\mathfrak{GlobHyp})$ with two embedded Cauchy surface $\tilde\Sigma\hookrightarrow\tilde M$ and $\tilde\Sigma'\hookrightarrow\tilde M$ as well as bcc. regions $U,V\subset M$ and $U',V'\subset M'$ such that, if we call $I^\pm$ the chronological future and past,
\begin{itemize}
\item there exist isometries $\psi^-$ between $I^-(\Sigma)$ and $I^-(\tilde\Sigma)$ as well as $\psi^+$ between $I^+(\Sigma')$ and $I^+(\tilde\Sigma')$,
\item $U',V'\subset I^+(\Sigma')$, $U'\subset D(\mathcal{O}')$ and $\mathcal{O}'\subset D(V')$, $D$ being the domain of dependence,
\item $U,V\subset I^-(\Sigma)$, $U,V^\perp\neq\emptyset$ and $\psi^-(U)\subset D(\psi^+(U'))$ as well as $\psi^+(V')\subset D(\psi^-(V))$.
\end{itemize}}

\noindent At a level of C$^*$-algebras, the above proposition has been used in the proof of the following facts:

{\proposition\label{def2} Let ${\rm W}:\mathfrak{GlobHyp}\to\mathfrak{Alg}$ be a covariant functor defining a locally covariant quantum field theory satisfying the time-slice axiom and let $S$ be the associated state space as per definition \ref{statespace}. Then
\begin{enumerate}
\item two $(M,g), (M',g')\in{\rm Obj}(\mathfrak{GlobHyp})$ with diffeomorphic Cauchy surfaces are mapped by ${\rm W}$ into isomorphic C$^*$-algebras,
\item for any bcc. region $\mathcal{O}'\subset M'$ with $\mathcal{O}^{\prime\perp}\neq\emptyset$, there exist bcc. open sets $U,V\subset M$ and a $*$-isomorphism $\alpha:{\rm W}(M',g')\to {\rm W}(M,g)$ such that $V^\perp\neq\emptyset$ and 
$${\rm W}(U,g|_U)\subset\alpha\left({\rm W}(\mathcal{O}',g'|_{\mathcal{O}'})\right)\subset {\rm W}(V,g|_V).$$
\end{enumerate}}

\noindent In turn this last proposition leads to the main statement of \cite{Sanders2}:

{\proposition\label{def3} Under the same assumptions of the previous proposition, the following statements hold true:
\begin{enumerate}
\item Let us consider $(M,g), (M',g')\in{\rm Obj}(\mathfrak{GlobHyp})$ with diffeomorphic Cauchy surfaces such that $\omega\in S(M,g)$ has the Reeh-Schlieder property. Then for any bcc. region $\mathcal{O}'\subset M'$ such that $\mathcal{O}^{\prime\perp}\neq\emptyset$, there exists a $*$-isomorphism $\alpha:W(M',g')\to W(M,g)$ such that $\omega'\doteq\alpha^*\omega$ has the Reeh-Schlieder property for $\mathcal{O}^\prime$. 
\item If the Cauchy surfaces are not compact, for any bcc. region $\mathcal{O}'_1\subset M'$, there exists a second bcc. region $\mathcal{O}'_2\subset\mathcal{O}_1^{\prime\perp}$ for which $\alpha^*\omega$ has the Reeh-Schlieder property.
\item If the locally covariant quantum field theory is causal in the sense of definition 2.1 in \cite{BFV}, then, given $(\mathcal{H}_{\omega'},\Pi_{\omega'},\Omega_{\omega'})$ -- the GNS triple of $\omega'$ --, it turns out that $\Omega_{\omega'}$ is cyclic and separating for $\overline{\pi_{\omega'}(W(\mathcal{O}',g'|_{\mathcal{O}'}))}^{\prime\prime}$. If the Cauchy surfaces are not compact, then $\omega'$ is also separating for all $\overline{\pi_{\omega'}(W(\mathcal{O}'_1,g'|_{\mathcal{O}'_1}))}^{\prime\prime}$, $\mathcal{O}'_1$ being an arbitrary bcc region in $M'$.
\end{enumerate}}

Notice that the above proposition deals only with the construction of a state which has the Reeh-Schlieder property only in suitably small neighbourhood. This is indeed the scenario we are interested in, but it is worthwhile to mention that a stronger result can be obtained: Let us consider a locally covariant quantum field theory ${\rm W}:\mathfrak{GlobHyp}\to\mathfrak{Alg}$, causal and satisfying the time-slice axiom with a locally quasi-equivalent state space -- see definition 2.4.3 in \cite{Sanders} for more details. Suppose that the latter is maximal, that is, for any state $\omega:{\rm W}(M,g)\to\mathbb{C}$ which is locally quasi-equivalent to a state in $S(M,g)$, $\omega$ lies in $S(M,g)$. Then, for any $(M,g), (M',g')\in\textrm{Obj}(\mathfrak{GlobHyp})$ with diffeomorphic and non compact Cauchy surfaces, $S(M',g')$ contains a full Reeh-Schlieder state if one such state exists in $S(M,g)$.

\subsection{Existence of a local Reeh-Schlieder state for higher spin field theories}
We can now use the analysis of the previous section to prove our main result. The line of reasoning can be summarized as follows: Since every globally hyperbolic spacetime $(M',g')$ is diffeomorphic to $\mathbb{R} \times\Sigma$, $\Sigma$ being a three-dimensional Cauchy surface, it is always possible to apply proposition \ref{def1}, \ref{def2} and \ref{def3} by fixing $(M,g)$ as an ultrastatic spacetime which is in turn diffeomorphic to $\mathbb{R}\times\Sigma$. In the latter case it was proven in \cite{Strohmaier} that, for a scalar or a Proca field, every quasifree and continuous state which is ground or KMS with respect to the timelike Killing vector field in $M$ has the Reeh-Schlieder property. Hence, since a real massive scalar field can be described as a locally covariant quantum field theory and since a ground state is always existent on a static spacetime \cite{Kay}, it is possible to combine the result of Strohmaier with proposition \ref{def3} to conclude the existence of a state for a real and massive scalar field theory on $(M',g')$ which satisfies the local Reeh-Schlieder property and, moreover, it is of Hadamard form.

Our goal is to generalize the above result as follows and we start from the case of Dirac fields.

{\proposition Let $(M',g',SM',\pi_S)\in{\rm Obj}(\mathfrak{SSpac})$ be given and let $\mathcal{O}\subset M'$ be any bcc. region with $\mathcal{O}^\perp\neq\emptyset$. Then, for ${\rm B}(M',g',SM',\pi_S)=\mathcal{A}(M',g')$ as in proposition \ref{DCQFT}, there exists always a Hadamard state $\omega'$ which has the Reeh-Schlieder property for $\mathcal{O}$. Furthermore, if $(\mathcal{H}_{\omega'},\pi_\omega{\omega'},\Omega_{\omega'})$ is the GNS triplet associated to $\omega'$, $\Omega_{\omega'}$ is both cyclic and separating for $\overline{\pi_{\omega'}(\mathcal{A}(\mathcal{O},g'|_{\mathcal{O}})})^{\prime\prime}$.}
\begin{proof}
In proposition \ref{DCQFT}, it was proven that a Dirac field can be described as a locally covariant quantum field theory which furthermore satisfies the time-slice axiom. Furthermore, a free Dirac field on an ultrastatic spacetime $(M,g)$ admits always a pure and quasifree state $\omega$ which is a ground state with respect to the dynamics induced at a C$^*$-algebra level by the timelike Killing field. This was first explicitly established in \cite{Fewster2} following theorem 2 in \cite{Araki} and \cite{Weinless}. Furthermore, on account of \cite{Sahlmann}, we know that such a state is of Hadamard form and, on account of \cite{Strohmaier}, that it has the Reeh-Schlieder property. Hence, we can invoke the deformation argument and propositions \ref{def2} and \ref{def3} to conclude the existence of a $*$-isomorphism $\alpha:{\rm B}(M',g',SM',\pi_S)\to {\rm B}(M,g,SM,\pi_S)$. This induces in turn a state $\omega'=\alpha^*\omega$ in $(M',g')$ which, on account of proposition \ref{def3}, has the sought properties. Furthermore, the very same deformation argument (see \cite{Sahlmann2}) entails also that $\omega'$ enjoys the Hadamard property. 
\end{proof}

\noindent For the massive spin $1$ field the outcome is not that different:

{\proposition Let $(M',g')\in{\rm Obj}(\mathfrak{GlobHyp})$ be given and let $\mathcal{O}\subset M'$ be any bcc. region with $\mathcal{O}^\perp\neq\emptyset$. Then there exists always a Hadamard state $\omega'$ for $\mathcal{W}_m(M',g')$ as in definition \ref{Weylmassive} which has the Reeh-Schlieder property for $\mathcal{O}$. Furthermore, if $(\mathcal{H}_{\omega'},\pi_{\omega'},\Omega_{\omega'})$ is the GNS triplet associated to $\omega'$, $\Omega_{\omega'}$ is both cyclic and separating for $\overline{\pi_{\omega'}(\mathcal{W}_m(\mathcal{O},g'|_{\mathcal{O}})})^{\prime\prime}$.}
\begin{proof}
In proposition \ref{LCQFTm} it was proven that a Proca field can be described as a locally covariant quantum field theory which, furthermore, satisfies the time-slice axiom. Hence we can apply the procedure depicted in the previous section by deforming the chosen globally hyperbolic spacetime $(M',g')$ into an ultrastatic one, $(M,g)$. On the latter the Weyl algebra for a massive spin $1$ field, $\mathcal{W}_m(M,g)$, admits a ground state with respect to the timelike Killing field \cite{Furlani}. On account of \cite{Sahlmann} we also know that such a state is of Hadamard form and, on account of \cite{Strohmaier} that it has the Reeh-Schlieder property. Hence, proposition \ref{def2} guarantees the existence of a $*$-isomorphism $\alpha:\mathcal{W}_m(M',g')\to\mathcal{W}_m(M,g)$ whereas proposition \ref{def3} asserts that $\omega'=\alpha^*\omega$ has the sought properties. Furthermore, the very same deformation argument (see \cite{Sahlmann2}) entails also that $\omega'$ enjoys the Hadamard property. 
\end{proof}

\noindent Slightly more complicated is instead the discussion for the vector potential. We remark that, in the case of globally hyperbolic spacetimes with trivial first de Rham cohomology group and compact Cauchy surface, one could shorten the argument by employing the results of \cite{Fewster}, section IV in particular. It holds:

{\proposition Let $(M',g')\in{\rm Obj}(\mathfrak{GlobHyp}_1)$ be given and let $\mathcal{O}\subset M$ be any bcc. region with $\mathcal{O}^\perp\neq\emptyset$. Then there exists always a Hadamard state $\omega'$ for $\mathcal{W}_0(M',g')$ as in definition \ref{Weylmassless} which has the Reeh-Schlieder property for $\mathcal{O}$. Furthermore, if $(\mathcal{H}_{\omega'},\pi_{\omega'},\Omega_{\omega'})$ is the GNS triplet associated to $\omega'$, $\Omega_{\omega'}$ is both cyclic and separating for $\overline{\pi_{\omega'}(\mathcal{W}_0(\mathcal{O},g'|_{\mathcal{O}})})^{\prime\prime}$.}
\begin{proof}
In proposition \ref{LCQFT0}, it was proven that the vector potential can be described as a locally covariant quantum field theory which satisfies, moreover, the time-slice axiom. Hence we can apply the procedure depicted in the previous section by deforming the chosen globally hyperbolic spacetime $(M',g')$ into an ultrastatic one, say $(M,g)$. Notice that, since the deformed manifold has the same topological structure of the former, it holds that $H^1(M)=\{0\}$. Unfortunately the existence of a state $\omega:\mathcal{W}_0(M,g)\to\bC$ which satisfies the Reeh-Schlieder property has never been explicitly established. Yet all ingredients to do it are already available in the literature and we can simply recollect them: First of all the existence of a quasi-free state $\omega$ for $\mathcal{W}_0(M,g)$ can be inferred from section 3.1 in \cite{Pfenning} where quantization via a Fock space is discussed. The main ingredients are the same as those needed for a scalar field: a weakly non degenerate symplectic form and a complex structure. Furthermore, since $(M,g)$ is ultrastatic, it possesses a timelike Killing field $\xi$ which generates a $1$-parameter group of isometries, say $g_t(\xi)$. This induces an action on each $f\in\Omega^1_0(M)$ via map composition and, since $g_t(\xi)$ is an isometry, it commutes with the action both of $\delta$ and of $d$. Hence, on account of definition \ref{equivrel}, both $\left[\Omega^1_{0,\delta}(M)\right]$ and $\mathcal{L}_0(M)$ are preserved by the natural action of $g_t(\xi)$. Furthermore, if one write \eqref{sympl} in a local coordinate system, it is immediate that $g_t(\xi)$ is also a symplectomorphism. Hence, according to these remarks, the state constructed in \cite{Pfenning} is a ground state and one could repeat almost slavishly the discussion in \cite{Strohmaier} to conclude that it also enjoys the Reeh-Schlieder property. On account of \cite{Sahlmann} we also know that such a state is of Hadamard form. Hence, proposition \ref{def2} guarantees the existence of a $*$-isomorphism $\alpha:\mathcal{W}_0(M',g')\to\mathcal{W}_0(M,g)$ which induces a state $\omega'=\alpha^*\omega$ with the sought properties since proposition \ref{def3} holds true. Furthermore, the very same deformation argument (see \cite{Sahlmann2}) entails also that $\omega'$ enjoys the Hadamard property. 
\end{proof}

\noindent To conclude the section, we would like to point out an additional feature of the deformation argument which is often neglected but descends almost automatically from the construction employed. Most notably, let us consider a globally hyperbolic spacetime $(M',g')$ and, up to an isometry, we can split $M'$ as $\bR\times\Sigma'$ and we can find a coordinate system $x^\mu=(t,x^i)$, $\mu=0,....,4$ and $i=1,...,3$ such that $g'_{\mu\nu}dx^\mu dx^\nu=-\beta dt^2+ h_{ij}dx^i dx^j$ where $\beta\in C^\infty(M,\bR^+)$ while $h$ is a smooth time-dependant Riemannian metric on the Cauchy surface $\Sigma'$. If there exists a complete spacelike Killing field $\xi$ for $g'$ whose integral curves lie, for each initial value, on a fixed Cauchy surface $\Sigma'$, it is possible to choose the ultrastatic spacetime $(M,g)$, which is diffeomorphic to $(M',g')$, in such a way that $\xi$ is also a spacelike complete Killing field for $g$. Furthermore each integral curve will lie entirely on a fixed Cauchy surface $\Sigma$ of $(M,g)$ which is diffeomorphic to $\Sigma'$. Hence, under these hypotheses and assuming the standard action of this additional isometry on the algebra of observables (the fermionic and the bosonic case yield the same result), the ground state $\omega$ we consider on the ultrastatic spacetime will be invariant under the action of the spacelike isometry. This invariance property will be preserved under the pull-back action of $\alpha$ introduced in proposition \ref{def2} and thus also $\omega'=\alpha^*\omega$ will be invariant under the action induced by $\xi$. 

\section{Conclusions}

We have proven that the Dirac and the Proca field as well as the vector potential admit even on a generic globally hyperbolic spacetime a state which enjoys the Reeh-Schlieder property at least on a suitable open region. To get to this result we also had to show that spin $1$ fields can be described as locally covariant quantum field theories regardless of the gauge freedom present in the massless case. From a mathematical point of view, this result guarantees that yet another property, first introduced on Minkowski background, admits a generalization on non trivial manifolds. From a physical point of view, we envisage instead at least two possible applications: As already commented in the introduction we expect that the states, we proved to exist, could be employed in the framework of warped convolutions where new field theory models are engineered by deforming free ones. The existence of a state which enjoys the Reeh-Schlieder property could allow to prove on a firmer ground that the resulting model is indeed non equivalent to the original one. From a completely different perspective, the outcome of this paper could be seen as a first step to bring our knowledge of spin $1$ fields on curved backgrounds on par with that of the other free fields. In particular the massless case has been often relegated to an ancillary role mostly due to the presence of the gauge freedom which makes the achievement of any result trickier to say the least. With this paper we want to start a series of analyses aiming to fill this gap and our next goal will be to extend the results of \cite{Fewster}. In this paper the very same deformation argument we employed was used to prove that, under certain topological restrictions on the Cauchy surface, a vector potential on a globally hyperbolic spacetime always admits at least one Hadamard state. We plan to show that the bulk-to-boundary procedure, first discussed in \cite{DMP}, can be used also for vector potential. To this avail the proof that such a field can be described in the language of general local covariance will play an important role and ultimately we will provide an explicit construction of Hadamard states on a large class of curved backgrounds.

\vspace*{9mm}

{\noindent\bf Acknowledgements}\\
C.D gratefully acknowledges financial support from the University of Pavia, from the project ``Stati Quantistici di Hadamard e radiazione di Hawking da buchi neri rotanti" funded by the GNFM-Indam and from the German Research Foundation DFG through the Emmy Noether Fellowship WO 1447/1-1. It is also gratefully acknowledged the hospitality of the II. Institut f\"ur Theoretische Physik - Universit\"at Hamburg, where part of this project was undertaken. C.D. is also in debt with Ko Sanders, Daniel Siemssen and Benjamin Lang for enlightening discussions.


\begin{thebibliography}{999}

\bibitem[Ar71]{Araki} H.~Araki, ``On Quasifree states of CAR and Bogoliubov Automorphsims'' Publ. RIMS Kyoto Univ. {\bf 6}, (1970/71) 385.

\bibitem[BaeFr09]{BF}
C. B\"ar, K. Fredenhagen (editors)
{\em ``Quantum Field Theory on Curved Spacetimes''} (2009) Springer.

\bibitem[BGP07]{BGP}
C.~B\"ar, N.~Ginoux and F.~Pf\"affle,
\emph{``Wave Equations on Lorentzian Manifolds and Quantization''}
(2007) European Mathematical Society.

\bibitem[BeSa03]{Bernal}
  A.~N.~Bernal, M.~Sanchez,
  {\em ``On Smooth Cauchy hypersurfaces and Geroch's splitting theorem,''}
  Commun.\ Math.\ Phys.\  {\bf 243 } (2003)  461-470,
  [gr-qc/0306108].

\bibitem[BeSa06]{Bernal2}
  A.~N.~Bernal, M.~Sanchez,
  {\em ``Further results on the smoothability of Cauchy hypersurfaces and Cauchy time functions,''}
  Lett.\ Math.\ Phys.\  {\bf 77 } (2006)  183-197,
  [gr-qc/0512095].

\bibitem[BFV03]{BFV}
  R.~Brunetti, K.~Fredenhagen and R.~Verch,
  {\em ``The generally covariant locality principle: A new paradigm for local quantum physics,''}
  Commun.\ Math.\ Phys.\  {\bf 237} (2003) 31, 
  [arXiv:math-ph/0112041].

\bibitem[BFM09]{Brunetti}
  R.~Brunetti, L.~Franceschini, V.~Moretti,
  {\em ``Topological features of massive bosons on two dimensional Einstein space-time. I: Spatial approach,''}
  Annales Henri Poincare {\bf 10 } (2009)  1027-1073.
  [arXiv:0812.0533 [gr-qc]].

\bibitem[BLS10]{BLS}
  D.~Buchholz, G.~Lechner, S.~J.~Summers,
  {\em ``Warped Convolutions, Rieffel Deformations and the Construction of Quantum Field Theories,''}
  [arXiv:1005.2656 [math-ph]], accepted on Comm. Math. Phys.

\bibitem[DMP06]{DMP}
  C.~Dappiaggi, V.~Moretti, N.~Pinamonti,
  {\em ``Rigorous steps towards holography in asymptotically flat spacetimes,''}
  Rev.\ Math.\ Phys.\  {\bf 18}, 349-416 (2006).
  [gr-qc/0506069].

\bibitem[DHP09]{DHP}
  C.~Dappiaggi, T.~P.~Hack and N.~Pinamonti,
  {\em ``The extended algebra of observables for Dirac fields and the trace anomaly
  of their stress-energy tensor,''}
  Rev.\ Math.\ Phys.\  {\bf 21} (2009) 1241,
  [arXiv:0904.0612 [math-ph]].

\bibitem[DLM10]{DLM}
  C.~Dappiaggi, G.~Lechner, E.~Morfa-Morales,
  {\em ``Deformations of quantum field theories on spacetimes with Killing vector fields,''}
  [arXiv:1006.3548 [math-ph]], accepted for publication on Comm. Math. Phys.

\bibitem[Dim80]{Dimock3}
J.~Dimock,
{\em ``Algebras of local observables on a manifold''},
Comm. Math. Phys. {\bf 77} (1980) 219.

\bibitem[Dim82]{Dimock}
J.~Dimock,
{\em ``Dirac Quantum Fields on a Manifold''},
Trans. Am. Math. Soc. {\bf 269} (1982) 133.

\bibitem[Dim92]{Dimock2}
  J.~Dimock,
  {\em ``Quantized electromagnetic field on a manifold,''}
  Rev.\ Math.\ Phys.\  {\bf 4}, 223 (1992).

\bibitem[FeVe02]{Fewster2}
  C.~J.~Fewster, R.~Verch,
  {\em ``A Quantum weak energy inequality for Dirac fields in curved space-time,''}
  Commun.\ Math.\ Phys.\  {\bf 225 }, (2002)  331,
  [math-ph/0105027].

\bibitem[FePf03]{Fewster}
  C.~J.~Fewster, M.~J.~Pfenning,
  {\em ``A Quantum weak energy inequality for spin one fields in curved space-time,''}
  J.\ Math.\ Phys.\  {\bf 44 }, (2003)  4480,
  [gr-qc/0303106].

\bibitem[FNW81]{FNW}
  S.~A.~Fulling, F.~J.~Narcowich, R.~M.~Wald,
  {\em ``Singularity Structure Of The Two Point Function In Quantum Field Theory In Curved Space-time. II''},
  Annals Phys.\  {\bf 136 }, (1981)  243.

\bibitem[Fre89]{Fredenhagen}
  K.~Fredenhagen,
  {\em ``Generalizations of the theory of superselection sectors,''}
  In {\it Palermo 1989, Proceedings, The algebraic theory of superselection sectors and field theory} (1989), 379.

\bibitem[Fur99]{Furlani}
  E.~P.~Furlani,
  {\em ``Quantization of massive vector fields in curved space-time,''}
  J.\ Math.\ Phys.\  {\bf 40}, (1999) 2611.

\bibitem[Ge68]{Geroch}
R.~Geroch,
{\em ``Spinor Structure of Space-Times in General Relativity. I''}
J. Math. Phys. {\bf 9}, (1968 1739.

\bibitem[Ge70]{Geroch2}
  R.~P.~Geroch,
  {\em ``Spinor Structure Of Space-Times In General Relativity. II,''}
  J.\ Math.\ Phys.\  {\bf 11}, (1970) 343.

\bibitem[Haag92]{Haag}
R.~Haag,
  {\em ``Local quantum physics: Fields, particles, algebras,''}
  Berlin, Germany: Springer (1992) 356 p. (Texts and monographs in physics).

\bibitem[Ha10]{Hack}
  T.~-P.~Hack,
  {\em ``On the Backreaction of Scalar and Spinor Quantum Fields in Curved Spacetimes,''}
Phd. Thesis - University of Hamburg, [arXiv:1008.1776 [gr-qc]].

\bibitem[Kay78]{Kay}
B.~Kay,
{\em ``Linear Spin-Zero Quantum Fields in External Gravitational and Scalar Fields - I''},
Comm. Math. Phys. {\bf 62}, (1978) 55.

\bibitem[KoNo63]{KNVOL1}
S.~Kobayashi and K.~Nomizu
{\em ``Foundations of differential geometry: Volume 1''}
(1963) Interscience Publisher.

\bibitem[Lang10]{Lang}
B.~Lang,
{\em ``Homologie und die Feldalgebra des quantisierten Maxwellfeldes''},
Diplomarbeit -  (2010) Universit\"at Freiburg.

\bibitem[Lee00]{Lee} 
J.~M.~Lee,
{\em ``Introduction to smooth manifolds''} (2000) Springer.

\bibitem[O'N83]{ONeill}
B.~O'Neill
{\em ``Semi-Riemannian Geometry''} (1983) Academic Press.

\bibitem[Pfe09]{Pfenning}
  M.~J.~Pfenning,
  {\em ``Quantization of the Maxwell field in curved spacetimes of arbitrary dimension,''}
  Class.\ Quant.\ Grav.\  {\bf 26 }, (2009) 135017, [arXiv:0902.4887 [math-ph]].

\bibitem[SaVe01]{Sahlmann2}
  H.~Sahlmann, R.~Verch,
  {\em ``Microlocal spectrum condition and Hadamard form for vector valued quantum fields in curved space-time,''}
  Rev.\ Math.\ Phys.\  {\bf 13}, 1203 (2001).
  [math-ph/0008029].

\bibitem[SaVe00]{Sahlmann}
  H.~Sahlmann, R.~Verch,
  {\em ``Passivity and microlocal spectrum condition,''}
  Commun.\ Math.\ Phys.\  {\bf 214 }, (2000)  705-731.
  [math-ph/0002021].

\bibitem[San08]{Sanders}
  J.~A.~Sanders,
{\em ``Aspects of locally covariant quantum field theory,''}
  Ph.D. Thesis (2008) University of York, arXiv:0809.4828 [math-ph].

\bibitem[San09]{Sanders2}
  K.~Sanders,
  {\em ``On the Reeh-Schlieder Property in Curved Spacetime,''}
  Commun.\ Math.\ Phys.\  {\bf 288 }, (2009)  271,
  [arXiv:0801.4676 [math-ph]].

\bibitem[San10]{Sanders3}
K.~Sanders.
{\em ``The locally covariant Dirac field''},
Rev. Math. Phys. {\bf 22}, (2010) 381,
[arXiv:0911.1304 [math-ph]].

\bibitem[SchVe08]{Schlemmer}
  J.~Schlemmer and R.~Verch,
  {\it ``Local Thermal Equilibrium States and Quantum Energy Inequalities,''}
  Annales Henri Poincare {\bf 9}, (2008) 945,
  [arXiv:0802.2151 [gr-qc]].

\bibitem[Stro00]{Strohmaier}
  A.~Strohmaier,
  {\em ``The Reeh-Schlieder property for quantum fields on stationary space-times,''}
  Commun.\ Math.\ Phys.\  {\bf 215}, (2000) 105,
  [math-ph/0002054].

\bibitem[Ver01]{Verch}
  R.~Verch,
  {\it ``A spin-statistics theorem for quantum fields on curved spacetime  manifolds
  in a generally covariant framework,''}
  Commun.\ Math.\ Phys.\  {\bf 223}, 261 (2001).

\bibitem[Wa84]{Wald}
R.~M.~Wald,
{\em ``General Relativity''} (1984) Chicago University Press.

\bibitem[Wein69]{Weinless}
M.~Weinless
{\em ``Existence and Uniqueness of the Vacuum for Linear Quantized Fields'',}
J. Funct. Anal. {\bf 4}, (1969) 350.
\end{thebibliography}
\end{document}